\documentclass[twocolumn,pra,superscriptaddress,floatfix]{revtex4}
\usepackage{amssymb}
\usepackage{graphicx}
\usepackage{bm}
\usepackage{amsmath}
\usepackage{amsfonts}

\begin{document}
\title{Stochastic dynamics of phase-slip trains and superconductive-resistive
switching in current-biased nanowires}
\author{David Pekker}
\affiliation{Department of Physics, Harvard University, 17 Oxford Street, Cambridge, MA 02138}
\affiliation{Department of Physics, University of Illinois at Urbana-Champaign, 1110 West
Green Street, Urbana, Illinois 61801, USA}
\author{Nayana Shah}
\affiliation{Department of Physics, University of Illinois at Urbana-Champaign, 1110 West
Green Street, Urbana, Illinois 61801, USA}
\author{Mitrabhanu Sahu}
\affiliation{Department of Physics, University of Illinois at Urbana-Champaign, 1110 West
Green Street, Urbana, Illinois 61801, USA}
\author{Alexey Bezryadin}
\affiliation{Department of Physics, University of Illinois at Urbana-Champaign, 1110 West
Green Street, Urbana, Illinois 61801, USA}
\author{Paul M.~Goldbart}
\affiliation{Department of Physics, University of Illinois at Urbana-Champaign, 1110 West
Green Street, Urbana, Illinois 61801, USA}

\begin{abstract}
Superconducting nanowires fabricated via carbon-nanotube-templating
can be used to realize and study quasi-one-dimensional
superconductors. However, measurement of the linear resistance of these
nanowires have been inconclusive in determining the low-temperature
behavior of phase-slip fluctuations, both quantal and thermal.  Thus,
we are motivated to study the nonlinear current-voltage characteristics in
current-biased nanowires and the stochastic dynamics of
superconductive-resistive switching, as a way of probing phase-slip
events. In particular, we address the question: Can a single phase-slip 
event occurring somewhere along the wire---during
which the order-parameter fluctuates to zero---induce switching, via the
local heating it causes? We explore this and related issues by
constructing a stochastic model for the time-evolution of the
temperature in a nanowire whose ends are maintained at a fixed
temperature. We derive the corresponding master equation as tool for
evaluating and analyzing the mean switching time at a given value of
current (smaller than the de-pairing critical current). The model indicates that
although, in general, several phase-slip events are necessary to
induce switching via a thermal runaway, there is indeed a regime of temperatures
and currents in which a single event is sufficient. We carry out
a detailed comparison of the results of the model with experimental 
measurements of the
distribution of switching currents, and provide an explanation for the
rather counter-intuitive broadening of the distribution width that is observed 
upon lowering the temperature. Moreover, we identify a regime in which the
experiments are probing individual phase-slip events, and thus offer a
way of unearthing and exploring the physics of nanoscale quantum 
tunneling of the
one-dimensional collective quantum field associated with the superconducting
order parameter.
\end{abstract}
\pacs{}
\maketitle

\section{Introduction}
\label{Sec:Introduction}
The fundamental process governing the collective physical properties
of quasi-one-dimensional superconducting systems is the phase-slip
process exhibited by the extended, complex-valued superconducting
order parameter field $\Psi(z)$, which depends on the position $z$
along the system.  In the course of a phase-slip process, the field
$\Psi(z)$, undergoes a transition from an initial (typically
metastable) supercurrent-carrying state $\Psi _{1}(z)$ to a final one
$\Psi_{2}(z)$.  In settings in which the voltage drop between the ends
of the system is externally controlled, these metastable states are
topologically distinct from one another: the total changes, $\int
dz\,d\Phi_{j}(z)/dz$, in their position-dependent phases
$\Phi_{j}(z)$, from one end of the system to the other, differ by
$2\pi$, and the supercurrents carried by these states differ,
too~\cite{FN:setting}.

In the absence of phase-slip processes, the order parameter field of
quasi-one-dimensional superconducting systems behaves reversibly,
i.e., energy stored as kinetic energy associated with supercurrent
remains undissipated.  If, however, phase-slip processes do occur, so
can dissipation, part of the coherent kinetic energy of superflow
being converted into incoherent motion, i.e., heat. For example, in a
voltage-controlled setting, current-reducing phase slips occur with a
higher frequency than current-increasing ones, leading to energy and
current dissipation and the notion of an intrinsic resistance, as
elucidated by Little~\cite{Little1967} and Langer and
Ambegaokar~\cite{LangerA1967}. Similarly, in a current-controlled
setting~\cite{McCumber1968}, the preferred sense of phase-slip
processes yields an average voltage consistent with a positive 
Joule-heating power. This is the sense in which phase-slip processes control
the collective properties of quasi-one-dimensional superconducting
systems: they constitute the building blocks via which one can
understand properties such as dissipation.

In principle, transitions in the state of the order-parameter field
can behave predominantly either classically or quantally, depending on
the temperature of the system. In the classical regime, being
metastable, the states are local minima of the classical free energy,
and the transitions between such states constitute thermal
fluctuations of the order-parameter field over the Arrhenius energy
barriers that separate these states. The study of the rates at which
such transitions occur, and their implications for collective 
charge-transport through superconducting nanowires, was initiated by
Little~\cite{Little1967}, and developed in detail, shortly thereafter,
by Langer and Ambegaokar~\cite{LangerA1967} and McCumber and
Halperin~\cite{McCumberH1970}. In the quantal regime, the transitions
between the states are quantum tunneling events, in which the entire,
extended, order-parameter field passes from one metastable state to
another through a classically forbidden region of field 
configurations~\cite{GrabertW1984, Giordano1990}. 

In this Paper we develop a theory of the kinetics of phase slips of
the superconducting order-parameter field in settings in which the
heat liberated or absorbed during these processes is not
instantaneously dissipated but, rather, leads to alterations in the
local temperature of the quasi-one-dimensional system and a resulting
flow of heat, which feed back to influence the phase-slip kinetics. As
discussed by Tinkham and co-workers~\cite{TinkhamFLM2003}, this
feedback leads to a {\it switching bistability\/} of the system involving a
pair of mesoscopic states: an essentially superconducting, low-voltage
state and a more highly resistive, high-voltage state. The rarity of
phase slips in the essentially superconducting state mean that very
little Joule heating takes place, which favors the persistence of this
state. However, the energy liberated by concentrated bursts of phase
slips can Joule-heat the system enough to weaken the
superconductivity, which enhances the likelihood of phase slips, 
and---via this feedback loop---lead to the essential destruction of the
superconductivity and the maintenance of the more strongly
Joule-heated, more highly resistive state.

Recent advances in sample preparation techniques have made possible
the fabrication and exploration of extraordinarily narrow nanowires.
These wires can be so fine, say $10\,\text{nm}$ wide or even less, 
that they bring within reach experimental
conditions in which heat liberated during small numbers of phase-slip
events---and perhaps even a {\it single\/} one---can have a dramatic
impact on the state of the wire, triggering the switching transition
from the superconducting to the resistive state.  Thus, by monitoring
the voltage between the ends of the wire one should be able to observe
macroscopic, or at least mesoscopic, consequences of an individual
phase slip, and hence investigate the properties of these nanoscale
building-blocks of the collective behavior of quasi-one-dimensional
superconductivity.  For example, one should be able to ascertain the
rate at which phase slips occur, and its dependence on temperature,
applied current, wire geometry and materials parameters.  One should
also be able to compare such rates with those suggested by theoretical
pictures in which the phase slips proceed primarily via thermal
activation over an energy barrier or via quantum tunneling through
one.  Hence, one should be able to move beyond the nanoscience that
observes the {\it structure\/} of nanomaterials or {\it
single-particle\/} phenomena within them, and progress towards a
nanoscience of collective processes.

The theory developed in this Paper aims to take a step beyond
Ref.~\cite{TinkhamFLM2003} by considering the {\it stochastic\/}
aspects of the phase-slip processes occurring in quasi-one-dimensional
superconducting systems, i.e., by allowing for sequences of
phase-slips that occur at {\it random\/} intervals of time, and exploring the
consequences of this stochasticity for the states exhibited by the
system. Our main focus will be on the implications of this underlying
stochasticity for the rate at which quasi-one-dimensional
superconducting systems undergo switching transitions from the
essentially superconducting state to the more highly resistive state,
as a function of the temperature and the current at which the system
is maintained. Experiments are commonly done in a mode in which the
current is not maintained at a fixed value but is, rather, repeatedly
ramped up at some fixed rate of increase, the current at which the switching
transition occurs being monitored, so as to produce a distribution of
switching currents, which depends on the temperature and the ramping
rate of the current~\cite{Sahu2008, Sahu2008b}. 

Motivation for work reported in this Paper comes from the kinds of experimental
investigations of superconductivity in nanowires touched upon in the
previous paragraph, and the concomitant need for a {\it road map\/} to
guide experimental investigations towards regimes of current and
temperature in which small numbers of phase-slip events, or even
single such events, induce switching transitions of essentially
superconducting wires to a highly resistive state.  Experiments
performed in this regime should provide access to the temperature- and
current-dependence of the rate at which individual phase-slip events
occur. A brief account of this work was reported in
Ref.~\cite{Shah2008}.

The Paper is structured as follows: In
Sec.~\ref{Sec:ExperimentalTarget} we describe the switching-current
experiments on hysteretic superconducting nanowires along with details of our
physical picture of superconducting-resistive switching. We
construct a stochastic model of the dynamics of the temperature in
superconducting nanowires wires in Sec.~\ref{Sec:Model}, and explore
its basic properties in Sec.~\ref{Sec:LangevinProperties}. Next, we
develop a formalism to address the the statistics of switching events
in Sec.~\ref{Sec:Formalism}, and use it to compute the switching rate
as a function of temperature and bias current in
Sec.~\ref{Sec:Results}, and, in turn, compare this rate to experiments
in Sec.~\ref{Sec:Comparison}. Finally, we present some concluding remarks
in Sec.~\ref{Sec:Conclusions}.

\section{Physical scenario for switching in current-biased nanowires}

\label{Sec:ExperimentalTarget}
\begin{figure}
\includegraphics[width=6cm]{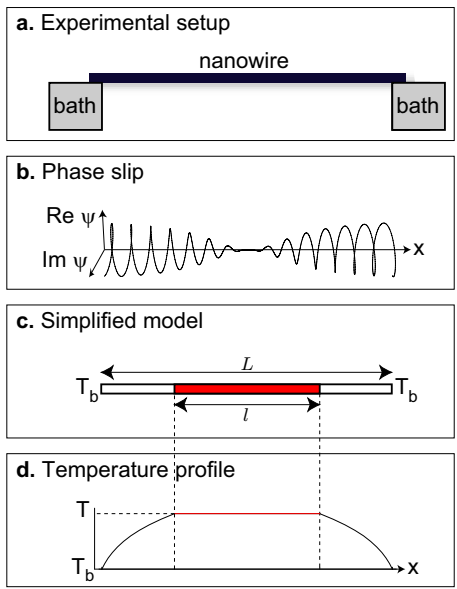}
\caption{Schematic showing {\bf a.}~the configuration of the
  free-standing nanowire supported by the superconducting leads, which
  act as a thermal bath; {\bf b.}~the attenuation of the order
  parameter $\psi(x)$ in the core of a phase slip; {\bf c.}~the
  simplified model of the wire: phase-slips occur exclusively in the
  central segment while the end segments carry the heat produced by
  phase-slips to the thermal baths; {\bf d.}~the temperature profile
  of the wire in the simplified model, with uniform temperature in the
  central segment and spatially varying temperature in the end segments. }
\label{Fig:Schematic}
\end{figure}

The ultranarrow wires that we consider in this Paper were fabricated
using molecular templating \cite{BezryadinLT2000, HopkinsPGB2005}.  By
using a solution containing long molecules such as carbon nanotubes
or DNA one can create a configuration in which a nanotube traverses a
trench, so as to form a bridge-like structure. One can then deposit a
layer of superconductor, such as MoGe or Nb, on top so
that the nanotube provides scaffolding on which to form a
superconducting nanostructure. In effect, one can thus fabricate a
set-up in which a free-standing superconducting nanowire is connected
at both of its ends to superconducting leads, as shown in
Fig.~\ref{Fig:Schematic}a. The diameter of the resulting nanowire can
be made smaller than the coherence length of the superconductor, and
the length of the wire sufficiently greater than the coherence length
so that the nanowire provides a realization of a quasi-one dimensional
superconductor in which the superconducting fluctuations are
effectively one dimensional. Through careful control, the wires
produced via molecular templating can be made amorphous and quite
homogeneous.  The resulting superconductor ends up being 
in the dirty limit (i.e., the electron mean-free-path is smaller than both
the coherence length and the penetration depth).

Upon lowering the temperature, the resistance of the nanostructure
(i.e., the leads and the nanowire) exhibits two drops: a sharp drop as
the leads become superconducting, and a second, much smoother
transition, corresponding to the onset of superconductivity in the
nanowire itself. The broad resistive transition of the nanowire can be
understood in terms of the occurrence of thermally activated
phase-slip fluctuations, and can be quantitatively fit in terms of
the \textrm{Langer-Ambegaokar-McCumber-Halperin} (LAMH) theory
\cite{LangerA1967, McCumber1968, McCumberH1970} by using the
transition temperature and the coherence length as fitting
parameters. However, the behavior of the resistance at very low
temperatures is not unambiguously established, either theoretically or
experimentally. On the one hand, time-dependent Ginzburg-Landau
theory, which forms the basis of the LAMH calculation, is not strictly
applicable in this regime of temperatures and, in addition, phase slip 
processes involving quantum
tunneling rather than thermal barrier crossing are expected to become
important in this regime. And on the other hand, the
value of the resistance can fall below the noise floor of the
experiment.

To overcome the difficulty in probing superconductivity in nanowires
at low temperatures associated with the smallness of the linear
resistance, we focus on experiments involving high bias-currents so
that they lie beyond the linear-response regime. In these experiments,
the current through the nanowire is ramped up and down in time, via a
triangular or sinusoidal modulation protocol. As the current is ramped
up, the state of the wire switches from superconductive to resistive
(i.e., normal), doing so at a value of the current that is smaller
than the de-pairing (i.e., equilibrium) critical current; and on
ramping the current down, the state gets re-trapped into a
superconductive state, but at a value of current smaller than the
current at which switching occurred. Hysteretic behavior such as this,
reflecting the underlying bistability of the superconducting nanowire
over a temperature-dependent interval of currents, was first reported
in Ref.~\cite{TinkhamFLM2003}. The experiments addressed in the
present paper~\cite{Sahu2008, Sahu2008b} go a step further, in that
they repeatedly ramp the current up and then down, for thousands of
cycles at each of a chosen set of temperatures, and thus generate
thousands of values of the switching and re-trapping currents at each
of these temperatures. These experiments find that the distribution of
re-trapping currents is very narrow and does not significantly change
with temperature. In contrast, the distribution of switching currents
is relatively broad, the mean and the width of the distribution
changing as the bath temperature---which is set by the leads---is
varied. The fact that even at a fixed temperature and current-sweep
protocol the switching current is statistically distributed and does
not have a sharp value is a reflection of (and therefore a window on)
the collective dynamics of the superconducting condensate in the
nanowire. The condensate is seen to be a fluctuating entity, evolving
stochastically in time and, at random instants, undergoing phase slip
events.  The goal of the present work is to understand the behavior of
these distributions of switching currents, and thereby gain insight
into the low-temperature rates at which thermal and quantum phase-slip
fluctuations occur. We now proceed to motivate the physical mechanism
for switching, and thus set the stage for the remainder of this Paper.

Distributions of switching currents were first studied in the context
of Josephson junctions, in work by by Fulton and 
Dunkleberger~\cite{FultonD1974}.
In particular, these researchers found that the width of the
distribution {\it decreased\/}, as the temperature was reduced.  As
will be discussed in detail in Sec.~\ref{Sec:Comparison}, in the
experiments on nanowires that we are considering~\cite{Sahu2008,
Sahu2008b}, this width is found to {\it increase\/} as the
temperature is reduced.
A second important difference is that the Josephson junctions that
show hysteresis are under-damped systems, whereas nanowires are expected to be
over-damped, as argued also in Ref.~\cite{TinkhamFLM2003}, i.e., a
single phase-slip event by itself is sufficient to cause switching
in hysteretic Josephson junctions but not in superconducting nanowires.
Experimentally, the observation of voltage tails~\cite{Rogachev,
AltomareCMHT2006, Sahu2008b}, i.e., small but non-zero voltages across
current-biased nanowires in the superconducting state, verifies the
occurrence of multiple phase slips prior to the switching event, and
indicates that the wire is in the over-damped regime.
The main consequence of these arguments is that while in the experiments
reported by Fulton and Dunkleberger the rate of switching is
essentially given by the rate $\Gamma$ at which individual phase-slips
occur, in the case of nanowires the switching rate is generically
found to be smaller than $\Gamma$.
Tinkham and co-workers~\cite{TinkhamFLM2003} have proposed a physical
mechanism that accounts for the fact that hysteresis in observed, in
spite of the over-damped dynamics of the wire.
According to this mechanism, the phase-slip fluctuations are resistive
but, because of the over-damping, they are not by themselves capable
of causing switching to a resistive (i.e., normal) state.
However, the resistance coming from the phase-slip fluctuations is
associated with Joule heating.  If this heating is not overcome
sufficiently rapidly (e.g., by conductive cooling) then it has the
effect of reducing the de-pairing current, ultimately to below the
applied current, thus causing switching to the highly resistive state.
Given that the wire is free-standing, heat can leave the wire only by
conducting it to the bath provided by the leads to which each of its 
ends is connected.
In the following sections, we discuss in more detail how bistability
and hysteresis come about within the framework of this physical
picture.  Along the way, it should become clear how, within this
picture, switching induced by multiple phase-slips can essentially be
interpreted in terms of the number of phase-slips needed to cause
a kind of thermal runaway instability of the superconducting state.

\section{Building a Model for Heating By Phase Slips}

\label{Sec:Model} 

The goal of this section is to construct a theoretical model of the
stochastic dynamics that leads to the switching of current-biased
nanowires from the superconductive to the resistive state. We begin by
reviewing the theory of the steady-state thermal hysteresis as set out
in Ref.~\cite{TinkhamFLM2003}. We continue by replacing the {\it
steady-state\/} heating of the wire by heating via {\it discrete\/}
stochastic phase-slips. The main result of this section is a
Langevin-type stochastic differential equation that describes the
dynamics of the temperature within the wire.

\subsection{Thermal hysteresis mechanism}

\label{Subsec:ThermalHysteresis} The thermal mechanism for hysteresis in
superconducting nanowires was originally proposed by Tinkham {\it et
al.\/}~\cite{TinkhamFLM2003}. The qualitative idea of this mechanism,
along with its relevance to experiments on superconducting nanowires,
was discussed in the previous section. In the present subsection we set up
the quantitative description of the mechanism by giving a brief
account of their work, and thus set the stage for our stochastic
extension of it. Their description rests on the premise that the
temperature of the wire is controlled by a competition between (i)~Joule
heating, and (ii)~cooling via the conduction of heat to the baths (leads). 
If ${\Theta}(x)$ is the temperature at position $x$ along the wire of
length $L$ and cross-sectional area $A$, then the power-per unit
length dissipated due to Joule heating at a bias current $I$
is taken to be
\begin{equation}
Q_{\text{source}}(x)=\frac{I^{2}R\big({\Theta}(x),I\big)}{AL}, \label{Q-cont}%
\end{equation}
where the function $R({\Theta}^{\prime})$ is to be understood as
the resistance of an entire wire held at a uniform temperature
${\Theta }(x)={\Theta}^{\prime}$. On the other hand, as the wire is
suspended in vacuum, the heat is almost exclusively dissipated through
its conduction from the wire to the superconducting leads that are
held at a temperature $T_{b}$ and which play the role of thermal
bath. The heating and cooling of the wire is described by the
corresponding static heat conduction equation:
\begin{align}
Q_{\text{source}}(x)  &  =-\partial_{x}\left[  K_{\text{s}}({\Theta
})\,\partial_{x}{\Theta}(x)\right] \label{HCE}\\
&  =-\frac{\partial K_{s}(\Theta)}{\partial\Theta}\left(  \partial_{x}%
\Theta\right)  ^{2}-K_{s}(\Theta)\partial_{x}^{2}\Theta, \label{Eq:HCE2}%
\end{align}
where $K_{\text{s}}({\Theta})$ is the thermal conductivity of the wire
(The first term on the right-hand side of Eq.~(\ref{Eq:HCE2}) was
absent in Ref.~\cite{TinkhamFLM2003}). This equation is supplemented
by the boundary conditions ${\Theta}(\pm L/2)=T_{\text{b}}$ at the
wire ends, $x=\pm L/2$, and we solve it numerically via the
corresponding discretized difference equation.

It was found in Ref.~\cite{TinkhamFLM2003} that Eq.~(\ref{Eq:HCE2})
yields two solutions for a certain range of $I\ $and
$T_{\text{b}}$. The nonlinear dependence of the resistance $R$ on
temperature, which is characteristic of a superconducting nanowire, is
at the root of this bistability. This bistability in turn furnishes
the mechanism for the thermal hysteresis in the $I$--$V$
characteristic; the two solutions correspond to the superconducting
(cold solution) and the resistive (hot solution) branches of the
hysteresis loop. To obtain the hysteresis loop at a given bath
temperature $T_{\text{b}}$, one begins by solving Eq.~(\ref{HCE}) to
obtain ${\Theta}(x)$ at a bias current sufficiently low such that the
equation yields only one solution. Next, by using the solution
${\Theta}(x)$ from the previous step to initialize the equation solver
for the next bias current step, the locally stable solution of
Eq.~(\ref{HCE}) is traced out as a function of $I$ by tuning the bias
current first up and then down. The $I$--$V$ loop is thus traced out
by calculating the voltage,
\begin{equation}
V=\int_{-L/2}^{L/2} dx\, I R\big({\Theta}(x),I\big),
\end{equation}
at each step.

The numerical analysis of Eq.~(\ref{HCE}) requires a knowledge of
$R\big(\Theta, I\big)$ and $K_{\text{s}}\big(\Theta\big)$, which serve
as input functions for the theory. A discussion of these input
functions and other parameters is given in
Appendix~\ref{App:Parameters}. In Ref.~\cite{TinkhamFLM2003}, the
linear-response resistance measured at $T_{\text{b}}=\Theta$ was used
for $R(\Theta, I)$.  However, $R(\Theta, I)$ depends also on the value
of the bias current $I$. Moreover, we find that by incorporating
deviations in $R(\Theta, I)$ from the linear-response regime, we are
also able to obtain a better fit with the experiments considered in
this Paper (see Ref.~\cite{Sahu2008b} and Appendix~\ref{App:Parameters}).

\subsection{Heating by discrete phase slip events: Derivation of Langevin
equation}

In the previous subsection, we described the static theory of thermal
hysteresis as was discussed in Ref.~\cite{TinkhamFLM2003} in the
context of experiments on MoGe nanowires. Let us now go one step
further and include dynamics by considering the time-dependent heat
diffusion equation:
\begin{equation}
C_{\text{v}}({\Theta})\,\partial_{t}{\Theta}(x,t)=\partial_{x}\left[
K_{\text{s}}({\Theta})\,\partial_{x}{\Theta}(x,t)\right]  +Q_{\text{source}},
\label{heat-cond-eq}%
\end{equation}
where the specific heat $C_{\text{v}}({\Theta})$ enters as an additional input
function. This differential equation can be derived in the standard way, by
using the continuity equation,
\begin{equation}
\nabla\cdot j_{Q}+\partial_{t}Q=Q_{\text{source}}%
\end{equation}
for the heat current,
\begin{equation}
j_{Q}\equiv-K_{s}(\Theta)\nabla\Theta,
\end{equation}
together with the energy density,
\begin{equation}
Q \equiv\int^{\Theta(x)}C_{v}(\Theta^{\prime})d\Theta^{\prime}.
\end{equation}

However, as long as we assume that the wire is heated by the source
term given by Eq.~(\ref{Q-cont}), the dynamic formulation turns out to
be inadequate for our purposes, as should become clear from our
analysis and results. Such a source term assumes that the wire is
being continually heated locally as a result of its resistivity
$R({\Theta}(x),I)A/L$ at any given position $x$ along the wire. Is
this assumption of continual Joule heating correct? To answer this
question we need to deconstruct the resistance and get to its root. In
Sec.~\ref{Sec:ExperimentalTarget} we have dwelt upon phase-slip
fluctuations in detail. There, we have emphasized the essential point
that it is the resistive phase-slip fluctuations that are responsible
for the characteristic resistance of a quasi-one dimensional
wire. Thus, one should consider the Joule heating as being caused by
individual, discrete phase-slip events.

Let us then explicitly consider discrete phase-slip events (labeled by $i$)
that take place one at a time at random instants of time $t_{i}$, and are
centered at random spatial locations $x_{i}$. By using the Josephson relation
\begin{equation}
\frac{d\phi}{dt}=\frac{2eV}{\hslash}=\frac{2\pi V}{\Phi_{0}} \label{JJ},
\end{equation}
relating the voltage pulse $V(t)$ to the rate of change of the end-to-end
phase difference across the wire $\phi$, we arrive at the work done on the
wire by a phase slip, viz.,
\begin{equation}
W_{\text{ps}}=\int dt\,IV=I\int_{0}^{2\pi}d\phi\,\frac{\hbar}{2e}=\Phi_{0}I,
\end{equation}
where $\Phi_{0}=h/2e$ is the superconducting flux quantum. Hence, a
single phase slip (or anti-phase slip), which corresponds to a
decrease (or increase) of $\phi$ by $2\pi$, will heat (or cool) the
wire by a ``quantum'' of energy $W_{\text{ps}}$. By using this result
we can now write down a time-dependent stochastic source term:
\begin{equation}
Q_{\text{source}}(x,t)\equiv\frac{W_{\text{ps}}}{A}\sum_{i}\sigma_{i} \,
F(x-x_{i})\, \delta(t-t_{i}), \label{Q-stoch}%
\end{equation}
where $F(x-x_{i})$ is a spatial form factor, of unit weight,
representing the relative spatial distribution of heat produced by the
$i^{\text{th}}$ phase-slip event, and $\sigma_{i}=\pm1$ for phase
(anti-phase) slips. The probability per unit time $\Gamma_{\pm}$ for
anti-phase (phase) slips to take place depends on the local
temperature ${\Theta}(x,t)$ and the current $I$.

Now, instead of using the continual Joule-heating source term,
Eq.~(\ref{Q-cont}), let us use the source term given by
Eq.~(\ref{Q-stoch}).  Instead of being a deterministic differential
equation, the heat diffusion equation~(\ref{heat-cond-eq}) becomes a
stochastic differential equation for $T(x,t)$. We thus have a Langevin
equation with stochasticity, in one space and one time dimension, with
a ``noise'' term that is characteristic of a jump process.

Let us pause to understand the connection between the two source
terms. By using the Josephson relation~(\ref{JJ}), we can express the
resistance as
\begin{equation}
R(\Theta,I)=\frac{V}{I}=\frac{1}{I}\frac{\Phi_{0}}{2\pi}\frac{d\phi}{dt}
=\frac{\Phi_{0}\, \Gamma(\Theta,I)}{I},
\end{equation}
and use it to rewrite the continual Joule-heating source term,
Eq.~(\ref{Q-cont}), as
\begin{equation}
Q_{\text{source}}=\frac{W_{\text{ps}}\,\Gamma(\Theta,I)}{A\,L},
\label{Eq:cont_stoch_connection}%
\end{equation}
where $\Gamma\equiv\Gamma_{-}-\Gamma_{+}$ is the net phase-slip rate
for the entire wire. Let us assume that a phase slip only affects its
local neighborhood, i.e., $F(x-x_{i}) \sim\delta(x-x_{i})$.  Then, if
we take the continuous time limit of Eq.~(\ref{Q-stoch}) by assuming
that the phase-slips are very frequent and that
$Q_{\text{ps}}\rightarrow0$, the two source terms would indeed become
equivalent (as can also be seen formally by taking the limit
$\Phi_{0}\rightarrow0$).

We now make a brief remark about the switching current. The static theory of
hysteresis that was discussed in the previous subsection has a single,
well-defined value of the switching current, which corresponds to the value of
the bias current at which the low-temperature (superconductive) solution
becomes unstable. On the other hand, we see from the theory discussed in the
present subsection that the randomness in $x_{i}$ and $t_{i}$ generates a
{\it stochasticity\/} in the switching process. The full implications of the
stochastic dynamical theory will be discussed in the following sections.

\subsection{Simplified model: Reduced Langevin equation}

In principle, one can proceed to study the physics of the stochastic switching
dynamics of a current-biased nanowire by using the dynamics of
Eq.~(\ref{heat-cond-eq}) together with stochastic source Eq.~(\ref{Q-stoch}),
both derived in the previous subsection. In practice, however, it is not easy
to solve the full Langevin equation with both spatial and temporal randomness.
In this subsection we derive a simplified model, and argue that it is capable
of capturing the physics essential for our purposes.

We concentrate on wires that are in the dirty limit, for which the
mean free path is much shorter than the coherence length, which is
shorter than the charge imbalance length required for carrier
thermalization, which itself is somewhat shorter than the nanowire
length $L$. In addition to restrictions on length-scales, we assume
that the time for a phase-slip ($\sim\tau_{\text{GL}}$) and the
quasi-particle thermalization time $\tau_{\text{E}}$ are both smaller
than the wire cooling time, i.e., the time it takes the heat deposited
in the middle of the wire by a phase-slip to diffuse out of the wire.

We will make a series of simplifications as follows:

\begin{enumerate}
\item Due to the presence of the superconducting leads at two ends, as
  well as edge effects, it is more likely that the phase-slip
  fluctuations in the wire are centered away from the wire edges. We
  thus assume that the source term is restricted to a region near the
  center of the wire.

\item We assume that the heating takes place within a central segment
  of length $l$, to which a uniform temperature $T$ is assigned. Note
  that the total length $L$ may be allowed to differ slightly from the
  geometric length of the wire, in order to compensate for the
  temperature gradients in the lead at the wire attachment points.

\item We assume that heat is conducted away through the end segments,
  each of which are of length ($L-l)/2$. As an additional
  simplification, we ignore the heat capacity of these end segments.

\item To simplify the problem further, we make use of the fact that
  the probability per unit time (i.e., the characteristic rate)
  $\Gamma_{+}$ for an anti-phase slip to take place is much smaller
  than the rate $\Gamma_{-}$ for a phase slip to take place, and we
  thus ignore the process of cooling by anti-phase slips. To account
  indirectly for their presence, we use a reduced rate
  $\Gamma\equiv\Gamma_{-}-\Gamma_{+}$ instead of $\Gamma_{-}$. This
  ensures that the discrete expression for $Q$ correctly reduces to
  the continual Joule-heating expression.
\end{enumerate}

With the simplified model defined above, the description of
superconducting nanowires reduces to a stochastic {\it ordinary\/}
differential equation for the time-evolution of the temperature $T$ of
the central segment:
\begin{equation}
\frac{dT}{dt}=-\alpha(T,T_{\text{b}})(T-T_{\text{b}})+\eta(T,I)\sum_{i}%
\delta(t-t_{i}). 
\label{SDE}
\end{equation}
This equation can be regarded as a spatially reduced Langevin-type
equation which is a counterpart to the full, spatially dependent
Langevin-type equation~(\ref{heat-cond-eq}) and, correspondingly, has
a spatially reduced version of the full source term~(\ref{Q-stoch}).
The second term on the RHS of Eq.~(\ref{SDE}) corresponds to
(stochastic) heating by phase slips, and the first to (deterministic)
cooling as a result of the conduction of heat from the central segment
to the external bath via the two end-segments. The
temperature-dependent cooling rate $\alpha(T,T_{\text{b}})$ is
obtained by comparing the heat currents through the end segments to
the thermal mass of the central segment, where the heat currents
through the end segments are found by solving the heat equation in the
end segments, subject to the boundary conditions that
$T(0)=T(L)=T_{\text{b}}$ and $T(\frac{L-l}{2})=T(L-\frac {L-l}{2})=T$.
The cooling rate may be expressed in terms of the integral
\begin{equation}
\alpha(T,T_{\text{b}})\equiv\frac{4}{l(L-l)C_{\text{v}}(T)}\frac
{1}{T-T_{\text{b}}}\int_{T_{\text{b}}}^{T}dT^{\prime}\,K_{\text{s}}(T^{\prime
}).
\end{equation}
If $T_{\text{i}}$ and $T_{\text{f}}$ are temperatures of the central segment
before and after a phase slip then we can express the temperature `impulse'
due to a phase slip, i.e., $T_{\text{f}}-T_{\text{i}}\equiv\eta(T_{\text{i}%
},I)\equiv\widetilde{\eta}(T_{\text{f}},I)$, as function of either
$T_{\text{i}}$ or $T_{\text{f}}$ (depending on the context) by using
\begin{equation}
A\text{ }l\int_{T_{\text{i}}}^{T_{\text{f}}}C_{\text{v}}(T^{\prime
})\,dT^{\prime}=Q_{\text{ps}}.
\end{equation}

To summarize, in this section we have derived a simplified model that
is described by the reduced Langevin equation~(\ref{SDE}). The central
assumption that we used to build this simplified model is that
phase-slips predominantly occur in the center of the nanowire, or at
least that their exact spatial locations are unimportant. This
assumption is appropriate for shorter nanowires, in which we do not
have several distinct locations along the wire at which a switching
event may nucleate. Specifically, if the wire length does not greatly
exceed the charge imbalance length (which itself is assumed to be much
larger than the coherence length) then, independent of where a
phase-slip occurs, the temperature profile in the wire after the phase
slip will be similar, and we are able to apply our simplified
time-only model.

\section{Basic properties of the Simplified Model: Bistability and switching}

\label{Sec:LangevinProperties} The goal of this section is to explain the
basic properties of the stochastic model that was formulated in the previous
section and is described by the reduced Langevin equation, Eq.~(\ref{SDE}).
First, we show explicitly that the competition between the heating and cooling
terms leads to the emergence of bistability. Next, we show that the stochastic
character of the heating term can lead to switching between the two metastable
states. Then, to characterize this switching we relate the lifetime of the
superconducting state to the Mean First Passage Time (MFPT) for the
temperature $T$ in the central segment to exceed a certain critical value
$T^{\ast}$.

\begin{figure}
\includegraphics[width=8cm]{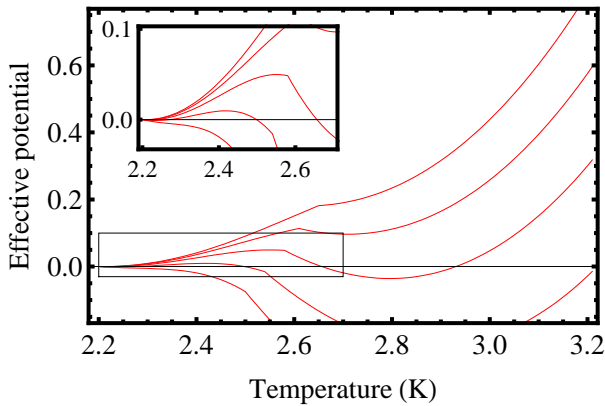}
\caption{Effective potentials for various bias currents ($0.175$,
  $0.195$, $0.215$, $0.235$, $0.255\,\mu\text{A}$) at fixed lead
  temperature $T_{\text{b}}=1.2\,\text{K}$~\cite{FN:parameters}. 
  The inset shows the details of the local
  maximum of the effective potential, and corresponds to an
  enlargement of the region indicated by the box in the main plot. }
\label{Fig:potential}
\end{figure}

We begin by reminding the reader that the theory of
Ref.~\cite{TinkhamFLM2003}, described by Eq.~(\ref{Eq:HCE2}) with the
source term given by Eq.~(\ref{Q-cont}), is the static
continual-heating version of the theory described by
Eq.~(\ref{heat-cond-eq}) with the source term given by
Eq.~(\ref{Q-stoch}). The connection is made evident via
Eq.~(\ref{Eq:cont_stoch_connection}). As has been shown at the end of
Subsection~\ref{Subsec:ThermalHysteresis}, for certain values of
$T_{\text{b}}$ and $I$ the continual-heating theory describes a
bistable system having a low-temperature superconducting branch and a
high-temperature resistive branch. The fluctuating theory described by
Eqs.~(\ref{heat-cond-eq}) and (\ref{Q-stoch}), in turn, allows for
processes that move the system between the two metastable states.

\begin{figure}
\includegraphics[width=8cm]{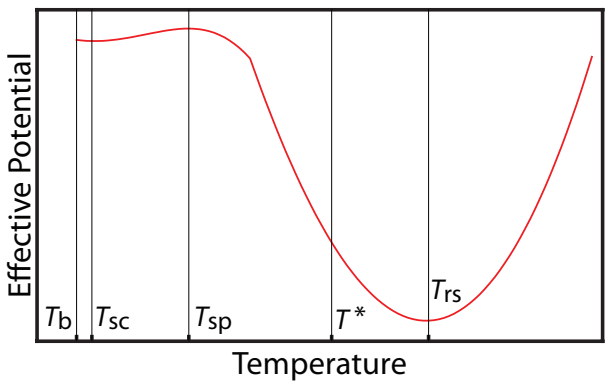}
\caption{Schematic depiction of the effective potential for the
  bistable case, showing the labeling scheme for the various
  temperatures: the bath temperature $T_{\text{b}}$,
  effective-potential minima corresponding to the superconducting
  state $T_{\text{sc}}$ and the resistive state $T_{\text{rs}}$, the
  local maximum of the effective potential $T_{\text{sp}}$, and the
  Mean First Passage Time point $T^{\ast}$.}%
\label{Fig:potentialLabels}%
\end{figure}

Analogously, there is a simplified continual-heating theory associated
with the spatially-reduced theory described by the reduced Langevin
equation~(\ref{SDE}). To obtain the continual-heating version of
Eq.~(\ref{SDE}), in analogy to Eq.~(\ref{Eq:cont_stoch_connection}),
we replace the term carrying the sum over the delta functions by the
phase slip rate $\Gamma(T,I)$. That is, if we ignore the discreteness
of phase-slips, the dynamics of the temperature is described by
\begin{align}
\frac{dT}{dt}=-\alpha(T,T_{\text{b}})\,(T-T_{\text{b}})+\eta(T,I)\,
\Gamma(T,I). \label{SCT}
\end{align}
We may think of this equation as describing the motion of an over-damped
\textquotedblleft particle\textquotedblright\ of position $T(t)$ at time $t$,
moving in the fictitious potential $U(T)$, i.e.,
\begin{align}
\frac{dT}{dt}=-\frac{\partial U(T)}{\partial T}.
\label{Eq:U}
\end{align}
In Fig.~\ref{Fig:potential} we have plotted the form of this potential
for several different values of the current at $T_{b}=1.2\,\text{K}$
for parameters corresponding to a typical nanowire. At low values of
the bias current, the fictitious potential $U(T)$ has only one
minimum, which corresponds to the superconducting, low-temperature
state with $T\approx T_{\text{b}}$. As the bias current is increased,
a second minimum, corresponding to the resistive state, develops at
higher temperatures, and the system becomes bistable. For the bistable
regime we label the temperatures of the two minima of the effective
potential by $T_{\text{sc}}$ and $T_{\text{rs}}$, and the temperature
of the local maximum that separates them by $T_{\text{sp}}$, as
depicted schematically in Fig.~\ref{Fig:potentialLabels}. Further
increase of the bias current results in the high-temperature minimum
gradually becoming deeper and the low-temperature minimum
shallower. Eventually, the low-temperature minimum disappears and only
the high-temperature minimum remains.

Consider a system biased such that it is bistable in the continual
heating description. If the fluctuations due to the discreteness of
phase-slips are weak, i.e., the temperature rise caused by an
individual phase slip is small compared to the temperature difference
between the two metastable minima, 
then the system highly likely to remain in whichever of the two minima
it started in. However, very rarely, the intrinsic fluctuations in the
times between phase-slips will drive the system from one local minimum
to the other.

\begin{figure}
\includegraphics[width=8cm]{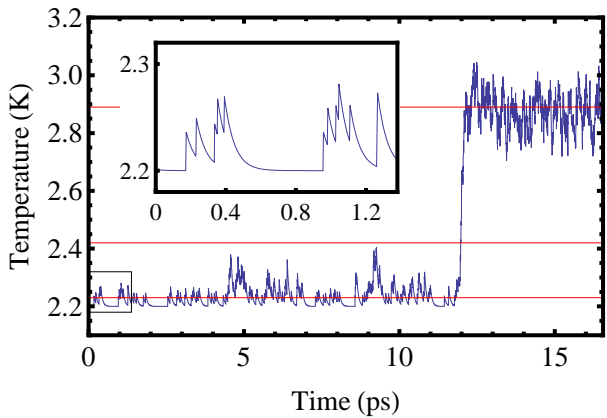}
\caption{Trace of a typical central segment temperature trajectory
  $T(t)$, with a switching event at
  $t\sim20\,\text{ps}$~\cite{FN:parameters}. Overlaid on the trace are
  red lines showing the temperatures $T_{\text{sc}}$ (lowest),
  $T_{\text{sp}}$ (middle), and $T_{\text{rs}}$ (highest)
  corresponding to the superconducting minimum, the saddle-point, and
  the resistive minimum of the effective potential. The inset shows a
  blow-up of the boxed region of the main figure. }
\label{Fig:Event}
\end{figure}

A picture of a switching event can be constructed by analyzing the
real-time dynamics of the Langevin equation. A typical trace of the
temperature of the central segment as a function of time $T(t)$,
evolving according to Eq.~(\ref{SDE}), is depicted in
Fig.~\ref{Fig:Event}. To obtain this trace we integrate
Eq.~(\ref{SDE}) forward in time, numerically, starting with the
initial condition $T(t=0)=T_{\text{b}}$. Phase slips correspond to
sharp rises of $T(t)$, and cooling to gradual declines of $T(t)$. The
two minima, $T_{\text{sc}}$ and $T_{\text{rs}}$, as well as the local
maximum $T_{\text{sp}}$ of the fictitious potential $U(T)$, are
indicated by the red lines. From the trace, it can be seen that the
temperature of the system starts by spending a long time in the
vicinity of $T_{\text{sc}}$, until a burst of phase slips pushes it
``over'' $T_{\text{sp}}$ and the temperature quickly progresses
towards the vicinity of $T_{\text{rs}}$.

The fundamental quantity of interest is the mean switching time,
i.e.,~the average time required for the wire to switch from being
superconductive to resistive. Assuming that the entire wire has
temperature $T=T_{\text{b}}$ when the current $I$ is turned on at time
$t=0$, we define a switching event as the first time at which $T$, the
temperature of the central segment of the wire, exceeds the
temperature $T^{\ast}$, where $T_{\text{sp}}\ <T^{\ast}\leq
T_{\text{rs}}$. With this definition, the mean switching time
corresponds to the Mean First Passage Time
$\tau(T_{\text{b}}\ \rightarrow T^{\ast})$ to go from the bath
temperature $T_{\text{b}}$ to the temperature $T^{\ast}$. In the case
of weak fluctuations, the problem is indeed a barrier crossing
problem, i.e., the system spends a long time in the vicinity of the
starting temperature $T_{\text{b}}\sim T_{\text{sc}}$, until a burst
of phase slips propels it over the barrier at $T_{\text{sc}}$. After
$T_{\text{sc}}$ is exceeded, the system moves quickly (compared to the
barrier-crossing time) towards $T_{\text{rs}}$. Therefore, the mean
switching time will only have a weak dependence on $T^{\ast}$,
provided $T^{\ast}$ is significantly higher than $T_{\text{sp}}$, as
this is the temperature range in which the system is moving relatively
quickly towards the high-temperature steady state.

\section{Formalism for addressing switching dynamics: Mean first passage time}
\label{Sec:Formalism}

In the previous section, we have shown that the Mean First Passage
Time (MFPT) for the temperature $T$ in the central segment to exceed a
critical value $T^{\ast}$ can be used to characterize the switching
from the superconducting to the resistive state. In this section, we
develop the tools for computing the MFPT in two steps. In the first
step, we derive the Master Equation associated with the Langevin
equation~\ref{SDE}. In the second step, we use the Master Equation to
obtain a delay-differential equation directly for the MFPT
$\tau(T)\left[\equiv\tau(T\rightarrow T^{\ast})\right]$ as a function
of the initial temperature $T$.

\subsection{Master Equation}

We now derive the Master Equation associated with the Langevin
equation~(\ref{SDE}). The Master Equation is a delay-differential
equation that describes the evolution of the probability density
$P(T,t)$ for the central segment of the nanowire to have temperature
$T$ at time $t$. Note that by the term ``delay'' we actually mean a
delay in temperature. That is, the evolution of $P(T,t)$ is non-local
in temperature but local in time. Before delving into the derivation,
we quote the result:
\begin{align}
\partial_{t}P(T,t)=  &  \partial_{T}\,[(T-T_{\text{b}})\,\alpha
(T)\,P(T,t)]-\Gamma(T)\,P(T,t)\nonumber\\
&  +\Gamma\big (T-\widetilde{\eta}(T)\big)\,P\big (T-\widetilde{\eta
}(T),t\big)\times\nonumber\\
&  \quad\quad\quad\quad\quad\quad\quad\quad\quad\times\big(1-\partial
_{T}\,\widetilde{\eta}(T)\big), \label{ME}%
\end{align}
where the first (i.e., the transport) term corresponds to the effect
of cooling, and the last two terms correspond to the effects of
heating. We remind the reader that, $\widetilde{\eta}(T)$ is defined
as the temperature change due to a phase slip, such that after the
phase slip the central segment temperature becomes $T$.

\begin{figure*}[ptb]
\includegraphics[width=11cm]{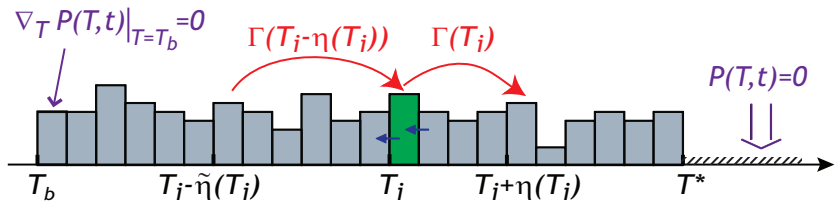}
\caption{Schematic diagram showing the probability $P(T_{j},t)$ to
  find the system in the temperature interval $(T_{j},\,T_{j}+\Delta T)$
  indicated by gray blocks. Processes involved in the rate of change
  of the probability in the $i-\text{th}$ interval (green block) are
  represented by blue arrows (cooling) and red arrows (heating by
  phase-slips). The boundary conditions at the ends of the temperature
  interval $(T_{\text{b}},\,T^{\ast})$ for computing the Mean First
  Passage Time are indicated in purple.}
\label{Fig:ME}
\end{figure*}

The probability distribution $P(T,t+dt)$ at temperature $T$ and time
$t+dt$ is related to $P(T,t)$ at an earlier time $t$ by the two
effects: (a) cooling, and (b) heating, as summarized in
Fig.~\ref{Fig:ME}. To understand these effects, we begin by
discretizing the temperature interval into equally wide slices $T_{i}$
of width $dT$, indexed by $i$.

To understand the effect of cooling, we consider the change in the
probability of finding the system in the $i$-th temperature slice,
$P(T_{i},t+dt)-P(T_{i} ,t)$. Due to cooling, some of the probability
in slice $i+1$ will move into the $i$-th slice at a rate
$(T_{i+1}-T_{b})\,\alpha(T_{i+1})/dT$.  Concurrently, some of the
probability slice $i$ will move into slice $i-1$ at a rate
$(T_{i}-T_{b})\,\alpha(T_{i})/dT$. These two processes are indicated
by the blue arrows in Fig.~\ref{Fig:ME}. By adding these two rates, we
find
\begin{align}
\partial_{t}P(T_{i},t)=\frac{1}{dT}  &  [(T_{i+1}-T_{b})\,\alpha
(T_{i+1})\,P(T_{i+1},t)\nonumber\\
&  -(T_{i}-T_{b})\,\alpha(T_{i})\,P(T_{i},t)]. \label{Eq:coolP}%
\end{align}
To find the continuum version of this equation we take the limit
$dT\rightarrow0$, thus arriving at
\[
\partial_{t}P(T,t)=\partial_{T}[(T-T_{b})\alpha(T)P(T,t)],
\]
where we have identified the definition of the $T$ derivative when taking this
limit on the right hand side of Eq.~(\ref{Eq:coolP}).

On the other hand, heating is caused by discrete phase-slips, which
occur at a temperature- and current-dependent rate
$\Gamma(T,I)$. Heating decreases the probability in the $i$-th slice,
$P(T_{i},t)$, at the rate $\Gamma (T_{i},I)\,P(T_{i},t)$, as the
probability is boosted to higher temperatures by phase-slips. On the
other hand, $P(T_{i},t)$ increases, as probability from lower
temperatures, $\widetilde{T}\approx T_{i}-\widetilde{\eta}(T_{i})$,
gets boosted to $T_{i}$ due to heating. These two heating processes
are indicated by the red arrows in Fig.~\ref{Fig:ME}. To compare these
two rates, we must take care of the fact that the boost is temperature
dependent, and thus there may be some `stretching' of the
corresponding temperature intervals before and after the
boost. Consider the temperature interval $(T_{i},\, T_{i}+dT)$ after the
boost. What is the corresponding temperature interval before the
boost?  Using the `un-boost function' $\widetilde{\eta}$, we find the
interval to be
\begin{align}
\big( T_{i}-\widetilde{\eta}(T_{i}),\,T_{i}+dT-\widetilde{\eta}(T_{i}+dT) \big).
\end{align}
Therefore, the width of the interval before the boost is approximately
$(1-\partial_{T_{i}}\,\widetilde{\eta}(T_{i}))dT$ and not $dT$. Thus, after
taking the continuum limit we find that the probability density $P(T)$
increases at the rate
\[
(1-\partial_{T}\,\widetilde{\eta}(T))\,f(T-\widetilde{\eta}%
(T))\,P(T-\widetilde{\eta}(T),t).
\]
Combining the rates of in- and out-flux of the probability-density due
to cooling and heating effects, we obtain the master equation given as
Eq.~(\ref{ME}).

\subsection{Mean First Passage Time}

The fundamental quantity that we want to compute is the Mean First
Passage Time (MFPT) $\tau(T_{\text{b}}\ \rightarrow T^{\text{*}})$. In
this subsection, we use the master equation~(\ref{ME}), as a starting
point for obtaining a delay-differential equation directly for
$\tau(T)\equiv \tau(T\rightarrow T^{\text{*}})$. We proceed using a
straightforward generalization of the standard procedure to systems
with jump processes (i.e., those that have master equations with delay
terms)~\cite{Gardiner, Kampen}.

We begin by supplementing the Master Equation with the boundary
conditions appropriate for computing the MFPT. When computing the
MFPT, we want to remove any element from the ensemble once its
temperature reaches one of the boundaries of the interval. Thus, we
would typically impose \textit{absorbing wall\/} boundary conditions
on both sides of the interval at $T_{\text{b}}$ and $T^{\ast}$, i.e.,
$P(T_{\text{b}},t)=P(T^{\ast},t=0)$. However, because we have a jump
process, which transfers probability-density from lower temperatures
to higher temperatures, we must instead impose the \textit{absorbing
  segment\/} boundary condition
\begin{align}
P(T,t)&=0, & \text{for }T>T^*,
\label{Eq:BCtop}
\end{align}
beyond the upper end of the interval, so as to capture systems in
which the temperature gets boosted beyond the upper end of the
interval by the jump process. The dynamics described by Eq.~(\ref{ME})
is also peculiar in another way. If we choose $T_{\text{b}}$ as the
temperature at the lower end of the interval then there are no
processes that can cause a passage through the lower end of the
interval, as can be seen from the Langevin equation~(\ref{SDE}). Thus,
at the lower end of the interval, instead of the absorbing wall
boundary condition, we must impose the \textit{no flux\/} boundary
condition
\begin{align}
\partial_{T_\text{b}} P(T_\text{b},t)&=0.
\label{Eq:BCbot}
\end{align}
These boundary conditions, (\ref{Eq:BCtop}) and (\ref{Eq:BCbot}), are
indicated (in purple) in Fig.~\ref{Fig:ME}. We note that the probability
density may be discontinuous at the upper boundary, due to the jump process.

With the boundary conditions in hand, we consider the integrated
probability density function
\begin{align}
G(T',t)=\int_{T_\text{b}}^{T^*} dT \, P(T,t | T',0).
\end{align}
Here, we have generalized from the probability $P(T,t)$ to the
conditional probability $P(T,t|T^{\prime},0)$ of finding the system to
be at temperature $T$ at time $t$, subject to the boundary conditions,
if it started out at temperature $T^{\prime}$ at time $0$. Then the
function $G(T^{\prime},t)$ measures the probability that a system that
started out at temperature $T^{\prime}$ has never left the temperature
interval $(T_{\text{b}},\,T^{\ast })$ while the time $t$ has
passed. In particular, the rate of first passages out of the interval
$(T_{\text{b}},\,T^{\ast})$ at time $t$ is given by
$-\partial_{t}G(T^{\prime},t)$. Therefore, the MFPT $\tau(T^{\prime})$
for a system that starts at temperature $T^{\prime}$ is given by
\begin{align}
\tau(T^{\prime})  &  =\int_{0}^{\infty}dt\,t\left(  -\partial_{t}G(T^{\prime
},t)\right) \\
&  =\int_{0}^{\infty}dt\,G(T^{\prime},t), \label{Eq:IntTau}%
\end{align}
where the surface term resulting from an integration by parts is assumed to be
zero, as all the \textquotedblleft particles\textquotedblright\ are assumed to
be able to leave the interval in the long-time limit.

Next, we obtain a differential equation for $\tau(x)$ by appropriately
integrating the backwards-in-time master equation, i.e., the equation
\begin{widetext}
\begin{align}
\partial_{t'} P(T,t|T',t')=(T'-T_\text{b}) \, \alpha(T') \, \partial_{T'} P(T,t|T',t')
+ \Gamma(T') \left[ P(T,t|T',t') - P(T,t|T'+\eta(T'),t')\right].
\label{Eq:BIMEfull}
\end{align}
\end{widetext}
Taking advantage of the fact that the present stochastic process
is homogeneous in time, we transfer the time-derivative on the left hand side
of Eq.~(\ref{Eq:BIMEfull}) to the $t$ variable from the $t^{\prime}$
variable:
\begin{align}
\partial_{t^{\prime}}P(T,t  |T^{\prime},t^{\prime})  &=+\partial
_{t^{\prime}}P(T,t-t^{\prime}|T^{\prime},0)\\
&  =-\partial_{t}P(T,t-t^{\prime}|T^{\prime},0).  \label{Eq:t_tPrime}%
\end{align}
By substituting Eq.~(\ref{Eq:t_tPrime}) into Eq.~(\ref{Eq:BIMEfull}) and
integrating both sides with respect to $T^{\prime}$ over the interval
$(T_{\text{b}},\,T^{\ast})$, we arrive at an equation for $G(T^{\prime},t)$:
\begin{align}
\partial_{t}G(T^{\prime},t)  &  =(T_{\text{b}}-T^{\prime})\,\alpha(T^{\prime
})\,\partial_{T^{\prime}}G(T^{\prime},t)\\
&  -\Gamma(T^{\prime})\left[  G(T^{\prime},t)-G(T^{\prime}+\eta(T^{\prime
}),t)\right]  .
\end{align}
Finally, we integrate over all times to obtain an explicit delay-differential
equation for the MFPT:
\begin{align}
 (T_{\text{b}}-T^{\prime})\,\alpha(T^{\prime})\,\partial_{T^{\prime}}
\tau(T^{\prime}) &  \label{Eq:tau}\\
-\Gamma(T^{\prime})&\left[  \tau(T^{\prime})-\tau(T^{\prime}+\eta(T^{\prime
}))\right]    =-1, \nonumber
\end{align}
where we have used Eq.~(\ref{Eq:IntTau}) to identify $\tau(T^{\prime})$ on the
left hand side, and the assumption that $P(T,t|T^{\prime},0)$ tends to zero in
the long-time limit on the right hand side.

The delay differential equation~(\ref{Eq:tau}), together with the
boundary conditions~(\ref{Eq:BCtop}) and (\ref{Eq:BCbot}), may be
conveniently solved numerically by using the shooting method. The key
to this method lies in taking advantage of the fact that in the
nonlocal term $\Gamma(T^{\prime})\,\tau(T^{\prime}+\eta(T^{\prime}))$,
the factor $\eta(T^{\prime})$ is always positive. Therefore, by
integrating from high temperatures to low temperatures, we can always
look up the value of the non local term from the region where
integration has already been carried out. We implement the shooting
procedure as follows: (1)~pick a value for $\tau(T^{\ast})$; (2)~shoot
towards lower temperatures to obtain $\partial_{T_{\text{b}}}%
\tau(T_{\text{b}})$; and (3)~adjust $\tau(T^{\ast})$ until the
boundary condition $\partial_{T_{\text{b}}}\,\tau(T_{\text{b}})=0$ is
satisfied.

\section{Switching behavior of superconducting nanowires: 
results and interpretations}
\label{Sec:Results}

\begin{figure}
\includegraphics[width=8cm]{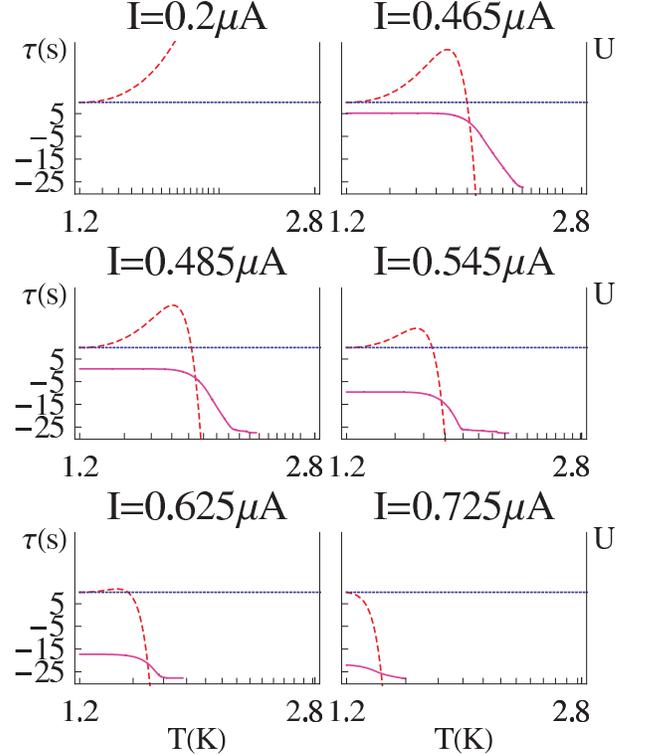}
\caption{Numerical solutions of the Mean First Passage Time
  Eq.~(\ref{Eq:tau}), as functions of the central segment temperature
  $T$, for bath temperature $T_{\text{b} }=1.2\,\text{K}$ and several
  values of the bias current between $0.2$ and $0.725\,\mu\text{A}$
  (solid pink)~\cite{FN:parameters}. For comparison, superposed on the
  MFPT plots are the effective potential that appears in
  Eq.~(\ref{Eq:U}) (dashed red lines). The solid blue lines correspond
  to the zero of the effective potential. The hash marks on the
  temperature axes correspond to the sequence of temperatures given by
  Eq.~(\ref{sequence}). }
\label{Fig:MFPT_and_potentials}
\end{figure}

\subsection{Properties of solutions of the Mean First Passage Time equation}

The Mean First Passage Time $\tau(T^{\prime}\rightarrow T^{\ast})$ for
a system that starts out at temperature $T^{\prime}$ to exceed the
temperature $T^{\ast}$ is described by the delay-differential
equation~(\ref{Eq:tau}), together with the boundary
conditions~(\ref{Eq:BCtop}, \ref{Eq:BCbot}). The mean switching time
corresponds to the MFPT $\tau(T_{\text{b}}\ \rightarrow T^{\ast})$. To
understand the solutions of the MFPT equations, we remind the reader
that $T^{\ast}$ must be chosen to be a temperature sufficiently far
above the saddle-point temperature $T_{\text{sp}}$ of the effective
potential (cf.~Fig.~\ref{Fig:potentialLabels}) that the MFPT has only
a weak dependence on $T^{\ast}$. Solutions of the MFPT equations for
several values of the bias current are plotted in
Fig.~\ref{Fig:MFPT_and_potentials}, along with the corresponding
effective potentials. The solutions have the following structure:

\begin{itemize}
\item {In the temperature interval $T_{\text{b}}<T<T_{\text{sp}}$ the
MFPT is largely independent of the temperature $T$.}
\item {For $T$ in the vicinity of $T_{\text{sp}}$ the MFPT drops sharply.}
\item {In the temperature interval $T_{\text{sp}}<T<(T^{\ast},T_{\text{rs}})$
the MFPT is again largely independent of $T$.}
\end{itemize}

The origin of this structure can be seen in the real-time dynamics
depicted in Fig.~\ref{Fig:Event}. Systems that start at temperatures
below the barrier temperature $T_{\text{sp}}$ in the effective
potential must diffuse over it, which is a very slow process. The MFPT
in the interval $T_{\text{b} }<T<T_{\text{sp}}$ is correspondingly
large. Furthermore, the MFPT is largely independent of the initial
temperature, as this interval is essentially ergodic, i.e., a system
that starts in this interval typically spends a lot of time exploring
the entire interval before leaving it. On the other hand, a system
that start at some temperature above the barrier in the effective
potential rolls down the potential gradient relatively quickly before
reaching $T^{\ast}$.  Therefore, in terms of temperatures increasing
from $T_{\text{b}}$, the MFPT starts out essentially constant over the
ergodic interval, and then drops sharply as $T$ crosses the barrier in
the effective potential, before finally flattening out at temperatures
higher than $T_{\text{sp}}$.

We note that it is numerically advantageous to set $T^{\ast}$ to be as
low as possible in the high-temperature (i.e., flat MFPT) regime, so
as to avoid instabilities in the numerical integration of
Eq.~(\ref{Eq:tau}).

\subsection{Number of phase slips in a thermal runaway train}

In this subsection, we consider the question of exactly how many
phase-slips it takes to form a runaway train. Let us start our
discussion by first turning off the (deterministic) {\it cooling\/}
term in the stochastic equation~(\ref{SDE}). If we now start with an
initial temperature $T_{0}$, then the sequence
\begin{equation}
T_{0},\,T_{0}+\eta(T_{0}),\,T_{0}+\eta(T_{0})+\eta(T_{0}+\eta(T_{0})),\,...
\label{sequence}%
\end{equation}
defines the discrete sequence of values that phase-slips would cause
$T$ to jump to, as marked on the horizontal axes in
Fig.~\ref{Fig:MFPT_and_potentials} for $T_{0}=T_{\text{b}}$. The
probability per unit time $\Gamma(T)$ to make a jump changes at each
step, and so does the size $\eta(T)$ of the jump, owing to their
explicit dependence on temperature.  On the other hand, if we turn off
the {\it heating\/} term then we would have a deterministic problem in
which $T$ would decay at a rate $\alpha(T)$, from its initial value
$T_{0}>T_{\text{b}}$ to the bath temperature $T_{\text{b}}$. It is the
competition between the discrete heating and continuous cooling
that makes for a rather rich stochastic problem.

The number of tick marks [see sequence~(\ref{sequence})] between
$T_{\text{b} }$ and $T^{\text{*}}$ (see
Fig.~\ref{Fig:MFPT_and_potentials}) is nothing but the of number
$N(T_{\text{b}},I)$ of phase-slip events required to raise the
temperature of the central segment from $T_{\text{b}}$ to
$T^{\text{*}}$ in the absence of cooling. Accordingly,
$N(T_{\text{b}},I)$ also provides an estimate of the number of
phase-slip events needed to overcome the potential barrier, if the
time-span of these events were insufficient to allow significant
cooling to occur. \textquotedblleft Thermal
runaway\textquotedblright---heating by rare sequences of
closely-spaced phase slips that overcome the potential
barrier---constitutes the mechanism of the
superconductive-to-resistive switching within our model. As the number
$N(T_{\text{b}},I)$ of phase-slips needed increases, the total number
of phase-slip events taking place before a switching event occurs, and
correspondingly the value of the mean switching time $\tau_{\text{s}%
}\ (T_{\text{b}},I)$, may indeed become quite large.

\begin{figure}
\includegraphics[width=8cm]{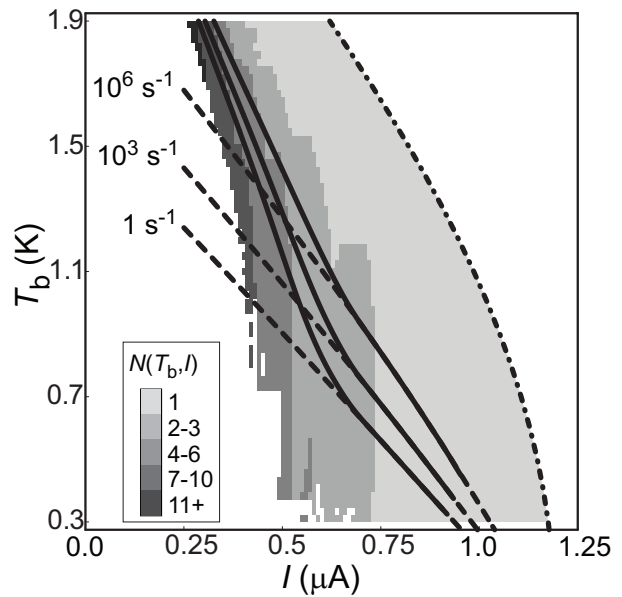}
\caption{Map of $N(T_{b},I)$, i.e., the number of consecutive phase
  slips required to overcome the barrier and induce the wire to
  switch, as a function of bias current $I$ and bath temperature
  $T_{b}$. The three solid lines are contours along which
  $\Gamma_\text{sw}^{-1}$ (the mean switching rate) is $1$, $10^{3}$,
  and $10^{6}$ $\text{s}^{-1}$. Similarly, the three dashed lines are
  contours along which the phase slip-rate is $1$, $10^{3}$, and
  $10^{6}$ $\text{s}^{-1}$. For comparison, the de-pairing critical
  current is indicated by the dashed-dotted line. }%
\label{Fig:Map}%
\end{figure}

The map of the number $N(T_{\text{b}},I)$ of phase slips needed over
the $I$-$T_{\text{b}}$ plane is presented in Fig.~\ref{Fig:Map}.  The
progression from smaller to larger $N(T_{\text{b}},I)$ corresponds to
the progression from lighter to darker shading. For clarity of
presentation we have grouped together a few values of
$N(T_{\text{b}},I)$. We see from the figure that, as the bias current
is increased, one traverses through regions of decreasing
$N(T_{\text{b}},I)$. It is important to observe that the typical value
of $N(T_{\text{b}},I)$ is only ten or fewer. The smallness of this
number highlights why going beyond continuous Joule heating to the
discrete phase-slip model is crucial for our analysis.

Remarkably, there is a region in the $I$-$T_{\text{b}}$ plane within which the
occurrence of \textit{just one phase-slip} is sufficient to cause the nanowire
to switch from the superconductive to the resistive state. We denote this this
region as the \textquotedblleft single phase-slip switching
regime\textquotedblright. A switching measurement in this range can in fact
provide a way of detecting and probing an \textit{individual phase-slip
fluctuation}.

\subsection{Mean switching time}

Let us begin by considering the single phase-slip switching regime identified
in the previous subsection. In this regime, the value of the mean switching
time is dictated purely by the probability for a phase-slip event to occur at
a given bath temperature $T_{\text{b}}$ and by the bias current $I$. The mean
switching rate is thus identical to the phase-slip rate (which is an input
quantity in our theory):
\begin{align}
\tau_{\text{s}}^{-1}  &  =\Gamma, &  &  \text{``single phase-slip switching
regime.''}%
\end{align}

\begin{figure}
\includegraphics[width=8cm]{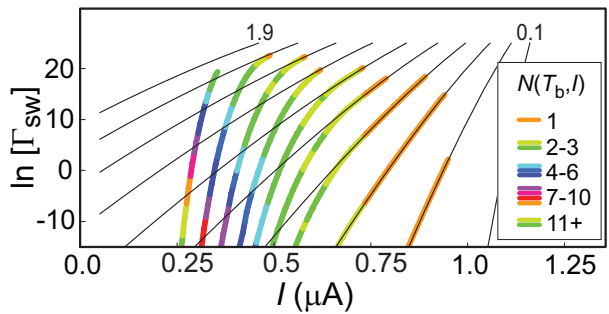}\caption{Logarithm
  of the mean switching rate, $\ln\Gamma_{\text{sw}}$, as a function
  of current for bath temperatures
  $T_{b}=0.1,\,0.3,\,0.5\,...,1.9\,\text{K}$.}
\label{Fig:MeanSwitchingTime}
\end{figure}

As we move beyond the \textquotedblleft single phase-slip switching
regime,\textquotedblright\ several phase-slip events become necessary
for switching. The mean switching rate $\tau_{\text{s}}^{-1}$ thus
begins to deviate from $\Gamma$ to a value below $\Gamma$, as can be
seen from Fig.~\ref{Fig:MeanSwitchingTime}.  For concreteness, we have
assumed that the phase-slip fluctuations are thermally activated, and
have used the form of $\Gamma$ as given by the LAMH theory for the
current-biased case. (The precise expression is provided in
Appendix~\ref{App:Parameters}.)\thinspace\ The deviation of
$\tau_{\text{s}}^{-1}$ from $\Gamma$ can be interpreted in terms of
the evolution in the number of phase slips $N(T_{\text{b}},I)$, which
was discussed in the previous subsection. To make the link between the
two more evident, we have color-coded the plots according to the value
of $N(T_{\text{b}},I)$.

An alternative representation of our results can be obtained by
studying the equal-value contours. A graphical representation of the
contour lines for a few values of $\tau_{\text{s}}^{-1}$ and $\Gamma$,
chosen in an experimentally accessible range, is provided in
Fig.~\ref{Fig:Map}. Whilst the spacing between the $\Gamma$ contour
lines decreases monotonically upon lowering $T_{\text{b}}$, the
spacing between the $\tau_{\text{s}}^{-1}$ lines can be seen to behave
non-monotonically. In the following subsection we examine a
consequence of this non-monotonicity.

\subsection{Switching-current distribution}

\begin{figure}
\includegraphics[width=8cm]{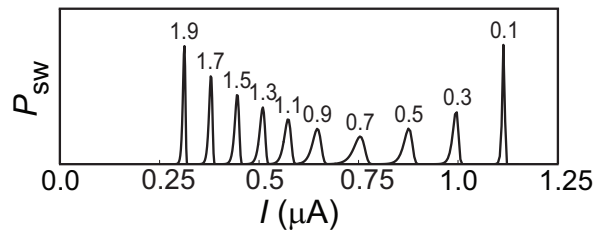}\caption{Switching rate
$P_{\text{sw}}$ as a function of current for bath temperatures $T_{b}%
=0.1,\,0.3,\,0.5\,...,1.9\,\text{K}$. The current sweep rate was set to
$58\mu\text{A}/\text{s}$.}%
\label{Fig:SwitchProb}%
\end{figure}

The mean switching time $\tau_{\text{s}}$ in bistable current-biased
systems can be measured directly by performing waiting-time
experiments.  Alternatively, $\tau_{\text{s}}$ can be extracted from
switching-current statistics~\cite{FultonD1974}. As described in
Section~\ref{Sec:ExperimentalTarget}, the switching-current
distribution can be generated via the repeated tracing of the $I$-$V$
characteristic, ramping the current up and down at some sweep rate
\[
\frac{dI}{dt}=\pm r.
\]
The sweep-rate-dependent probability $P(I,T_{\text{b}};r)dI$ for the
event of switching (from the superconductive to the resistive branch)
to take while the current is in the range $I$ to $I+dI$ is explicitly
related to the mean switching time $\tau_{\text{s}}$ via the relation
\begin{align}
P(T_{\text{b}},I;r)dI  &  =\left( \tau_{\text{s}}^{-1}(T_{\text{b}}%
,I)\frac{dI}{r}\right)  \times\nonumber\\
&  \quad\quad\times\left(  1-\int_{0}^{I}P(T_{\text{b}},I^{\prime
};r)\text{ }dI^{\prime}\right)  . \label{switch-dist}%
\end{align}
The term in the first pair of parentheses corresponds to the
probability for switching to happen within the ramp time, whilst the
term in the second pair of parentheses corresponds to the probability
that the wire has not already switched before reaching the bias
current $I$. By using Eq.~(\ref{switch-dist}) we obtain the
distribution of switching currents in superconducting nanowires in
terms of the theory presented in the present paper. The plots
illustrated in Fig.~\ref{Fig:SwitchProb} are simply the translation of
the mean switching-rate plots shown in
Fig.~\ref{Fig:MeanSwitchingTime}, to the switching-current
distribution for a chosen value of $r$.

Upon raising $T_{\text{b}}$, one would na{\"\i}vely expect the
distribution of the switching currents to become broader for a model
involving thermally activated phase slips. Such broadening of the
distribution is indeed obtained up to a crossover temperature scale
$T_{\text{b}}^{\text{cr}}(r)$ (i.e.,~the temperature below which,
loosely speaking, switching is induced by single phase
slips). However, on continuing to raise $T_{\text{b}}$, but now
through temperatures above $T_{\text{b}}^{\text{cr}}(r)$, the
distribution-width shows a seemingly anomalous decrease. Based on the
discussion of the previous subsection, this can be understood as a
manifestation of the now-decreasing spacing between the
$\tau_{\text{s}}$ contour lines in the $I$-$T_{\text{b}}$ space.

This striking behavior above $T_{\text{b}}^{\text{cr}}(r)$ may be
understood via the following reasoning: the larger the typical number
of phase-slips in sequences that induce superconductive-to-resistive
thermal runaway, the smaller the stochasticity in the switching process
and, hence, the sharper the distribution of switching currents. This
non-monotonicity in the temperature dependence of the width of the
switching-current distribution, along with the existence of a regime
in which a single phase-slip event can be probed, are the two key
predictions of our theory. In the following section, we proceed to
carry out a detailed comparison of between this theory and the
recently performed experiments discussed in
Sec.~\ref{Sec:ExperimentalTarget}.

\section{Comparison with experiments}
\label{Sec:Comparison}
In this section, we compare results from the experiments described in
Section~\ref{Sec:ExperimentalTarget} with predictions of our theory
presented in Sections~\ref{Sec:Model}-\ref{Sec:Results}. We show that
our theory is both qualitatively and quantitatively consistent with
experimental observations.  The main implications of this comparison
are that: (1)~the switching-current distribution-width does indeed
increase as the temperature is decreased; (2)~there is a single
phase-slip-to-switch regime at low temperatures; and (3)~thermally
activated phase-slips, alone, are insufficient to fit the dependence
of the mean switching time on the bias current at low temperatures.
This suggests that one should include the effects of Quantum
Phase-Slips (QPS)~\cite{Giordano1990, BezryadinLT2000}; upon including
quantum phase-slips phenomenologically, we obtain good fits to the
experimental data in the low-temperature regime as well. For the
purposes of this comparison we use the data from a representative
superconducting nanowire; data for more samples may be found in
Refs.~\cite{Sahu2008, Sahu2008b}. This section is structured as
follows. To establish the validity of the thermal hysteresis model, we
begin by analyzing the $I$-$V$ hysteresis loops. Next, we
qualitatively analyze the experimentally measured switching-current
distribution. We continue with a quantitative analysis of the
experimental data on the mean switching rate. Finally, we look at the
implications of the quantitative analysis, including identifying the
single phase-slip-to-switch regime and the scenarios for quantum
phase-slips in the low temperature behavior.

\subsection{I-V hysteresis loops}

\begin{figure}
\includegraphics[height=13cm]{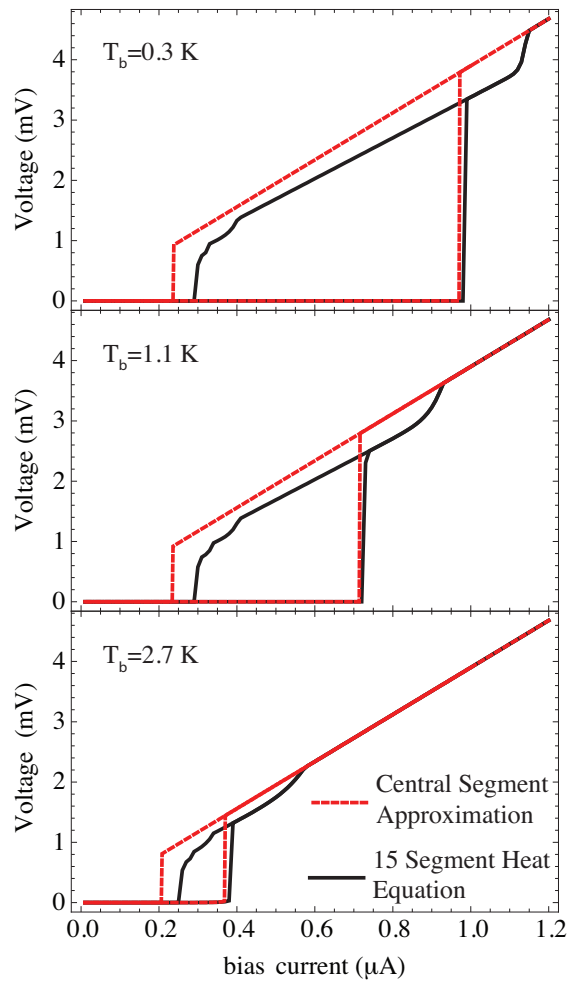}
\caption{Comparison of $I$-$V$ hysteresis loops computed using central
segment approximation and the full heat equations. For the case of the
heat equation the wire was split into 15 segments. The steps and
upturns in the high voltage branch of the segmented solution
correspond to various segments becoming superconducting, and would
disappear in the continuum limit. Heat conductivity, phase-slip rate,
and geometrical parameters used are listed in
Table~\ref{Table:Parameters}.}%
\label{Fig:Hysteresis}%
\end{figure}

We begin our analysis by comparing the qualitative features of the
experimentally measured and theoretically computed current-voltage
characteristics. Experimentally, it is found that at high temperatures
there is no hysteresis. As $T_{\text{b}}$ is lowered, a hysteresis
loop gradually opens up. Next, as the temperature is lowered even
further, the switching current (i.e., the bias current at the
superconducting-resistive transition) grows gradually, whilst the
re-trapping current (i.e., the bias current at the
resistive-superconducting transition) remains almost unchanged. This
behavior is consistent with the experimental observations and theory
of Ref.~\cite{TinkhamFLM2003}, where it is also qualitatively
explained as follows. Switching is controlled by the properties of the
low-temperature (i.e., superconducting-like) solution of
Eq.~(\ref{HCE}), thus switching depends strongly on the temperature of
the bath $T_{\text{b}}$. On the other hand, re-trapping is largely a
property of the hotter (i.e., resistive) state, and thus has only a
weak dependence on $T_{\text{b}}$.

Typical $I$-$V$ curves obtained from the central segment model [e.g.,
  stationary solutions of Eq.~(\ref{SCT})], as well as those obtained
from solving the heat equation, are shown in Fig.~\ref{Fig:Hysteresis}
for several bath temperatures $T_{\text{b}}$. Following
Ref.~\cite{TinkhamFLM2003}, the solutions of the heat equation were
obtained from a spatially discretized version of Eq.~(\ref{HCE}). In
both cases, the heat conductivity and the phase-slip rate were
obtained from Eqs.~(\ref{Eq:Cv}) and~(\ref{Eq:f}), respectively. The
theoretical curves both qualitatively and quantitatively reproduce the
features seen in
experiments~\cite{TinkhamFLM2003,Sahu2008,Sahu2008b}. We take a moment
to point out that in fitting the experimental data it is important to
take into account the nonlinear dependence of the phase-slip rate on
the bias current. Finally, we point out that making the central
segment approximation has little effect on the switching current found
in the hysteresis loops (for typical wires used in experiments). This
fact supports the validity of the central segment approximation for
modeling switching phenomena.

\subsection{Switching current distributions}

\begin{figure}
\includegraphics[width=7cm]{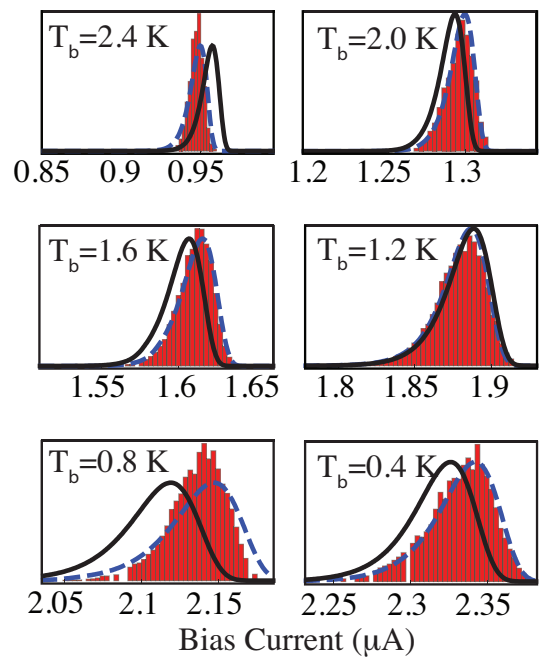}
\caption{Experimental data (red bar-charts) and theoretical fits
  (black lines) for the switching-current distributions for various
  bath temperatures. To make the comparisons between the shapes of
  distributions easier, we also show theoretical curves shifted so
  that their means coincide with the experimental curves (blue dashed
  lines).  The fitting parameters used are listed in
  Table~\ref{Table:Parameters}.}
\label{Fig:distributions}
\end{figure}

\begin{figure}
\includegraphics[width=8cm]{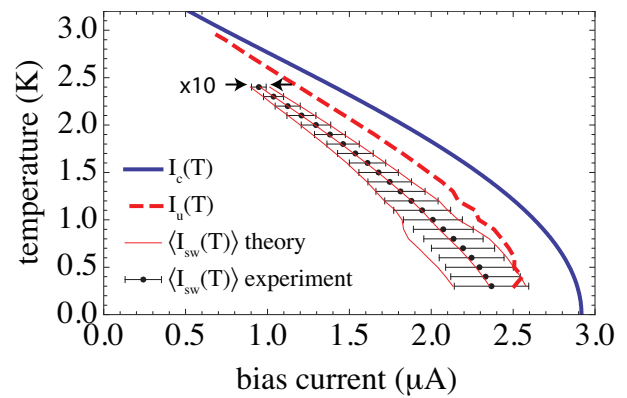}
\caption{Hierarchy of critical and switching currents. All theoretical
  curves were produced using the parameters obtained from fitting
  experimental data. The highest current scale is the de-pairing
  critical current $I_{\text{c}}(T)$. Next, is $I_{\text{u} }(T)$, the
  scale for linear instability due to overheating, as described by
  Eq.~(\ref{SCT}). Finally, comes the actual switching current
  $I_{\text{sw} }(T)$. The error bars on the switching current
  correspond to the $10\times$ the standard deviation of the switching
  current distribution (the scale of the standard deviation was
  exaggerated to make it easier to see; the sweep rate was set to
  $58\,\mu\text{A}/\text{s}$). The fitting parameters used are listed
  in Table~\ref{Table:Parameters}.}
\label{Fig:Ic}
\end{figure}

In the experiments, every time an I-V characteristic is measured by
sweeping the bias current up and down, switching occurs at a distinct
value of the bias current. By repeatedly measuring the hysteresis loop
at a fixed bath temperature $T_{\text{b}}$ and current sweep rate
$dI/dt$, one can obtain the distribution of switching currents
$P(I_{\text{sw}},T_{\text{b}},dI/dt)$.  Typical
$P(I_{\text{sw}},T_{\text{b}},dI/dt)$ distributions, obtained
experimentally, are shown in Fig.~\ref{Fig:distributions}. For
completeness, we also show the corresponding theoretical fits, which
we shall describe in detail in the next subsection. For a given
$T_{\text{b}}$, switching events tend, in general, to occur at lower
bias currents than the switching current found in the thermodynamic
stability analysis of Ref.~\cite{TinkhamFLM2003}. The reason for this
premature switching at bias currents that are lower than the stability
analysis indicates is, of course, thermally activated barrier crossing
in the form of a phase-slip bursts. In Fig.~\ref{Fig:Ic} we plot the
mean and the standard deviation of the switching-current distributions
measured experimentally, as well as those obtained from theoretical
fits of the simplified model. By using the tuning parameters obtained
from the fits, we also plot the theoretical de-pairing critical
current and the critical current from the stability analysis of the
simplified model Eq.~(\ref{SCT}).

\begin{figure}
\includegraphics[width=8cm]{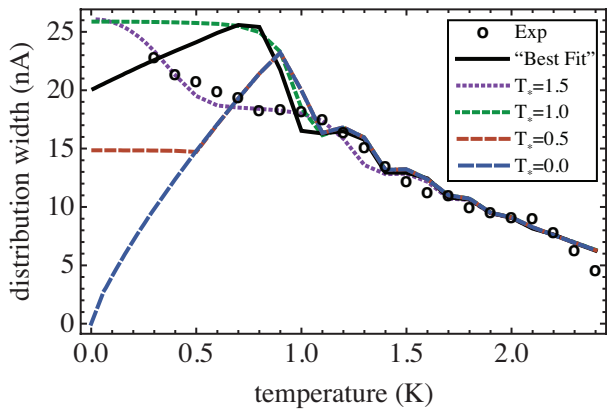}
\caption{Standard deviation of the switching-current distribution as a
  function of temperature. Comparison of experimental data for a
  typical sample (circles) with various QPS scenarios labeled by
  $T_{\ast}$. The ``best fit'' curve was obtained by using the
  parameters given in Table~\ref{Table:Parameters}, where the fit was
  optimized to simultaneously capture the temperature-dependence of
  mean switching current and the standard deviation of the switching
  current distribution. }
\label{Fig:distributionsVsT}
\end{figure}

As described in the introduction, one would typically expect the
standard deviation of the switching-current distribution to
\textit{decrease\/} with decreasing temperature, as thermal
fluctuations become suppressed. Such narrowing of the
switching-current distribution is expected to continue with cooling,
until the temperature becomes sufficiently low such that
\textit{quantum\/} phase slips are the main drivers of the switching,
at which point the narrowing is expected to come to a
halt. Qualitatively, this would indeed be the case if switching was
always triggered by a single phase-slip. However, our theory, predicts
that the situation is more complicated, because the mean switching
time, and hence the width of the switching-current distribution, is
controlled by a competition between the phase-slip rate and the number
of consecutive phase-slips needed to induce switching, as described in
the previous section. Thus, qualitatively, we expect the opposite
behavior at higher temperatures. That is, in the regime of thermally
activated phase slips and at temperatures above the single
phase-slip-to-switch regime, the width of the switching-current
distribution should \textit{increase\/} with decreasing
temperature. This counter-intuitive broadening of the switching
current distributions with decreasing temperature is indeed observed
experimentally, as shown in Fig.~\ref{Fig:distributionsVsT}.

\subsection{Mean switching rate}

\begin{figure}
\includegraphics[width=8cm]{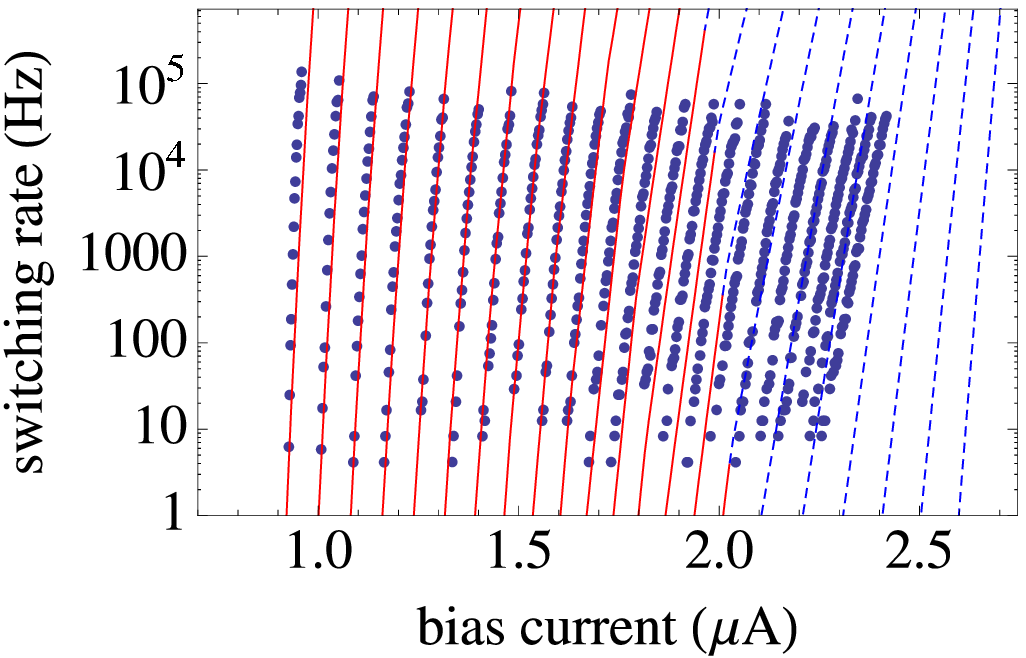} \newline%
\newline%
\includegraphics[width=8cm]{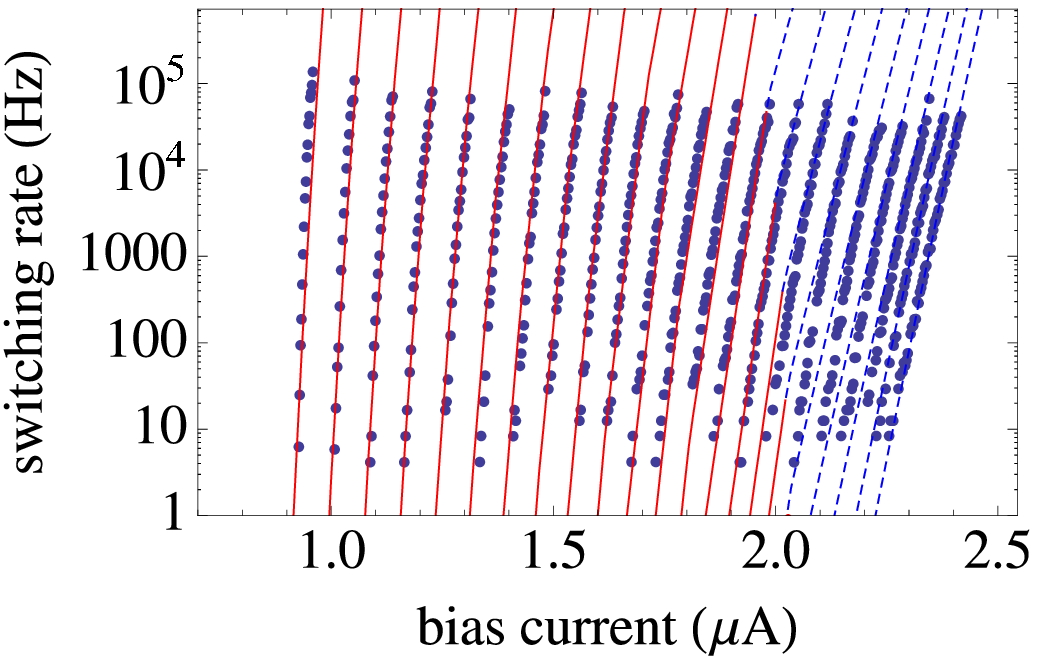}
\caption{Comparison of theoretical fits (lines) to the experimental
  (blue dots) Mean Switching Rates. The top panel shows fits with
  thermally activated phase slips only, whilst the bottom panel
  includes both thermally activated and quantum phase slips. The solid
  red lines correspond to multiple-phase-slips-to-switch regime,
  whilst the dashed blue lines correspond to
  multiple-phase-slips-to-switch regime. The fitting parameters used
  are listed in Table~\ref{Table:Parameters}.}%
\label{Fig:MSR}%
\end{figure}

As the switching-current distribution depends on the bias-current
sweep-rate, in order to quantitatively compare our theory with
experimental data, we focus on the mean switching rate, which is
related to the switching current distribution via
Eq.~(\ref{switch-dist}). The experimentally obtained mean switching
rate $\tau^{-1}(I,T_{\text{b}})$ for a typical sample, along with
theoretical fits, are plotted as a function of the bias current $I$
for different values of the bath temperature $T_{\text{b}}$ in
Fig.~\ref{Fig:MSR}.  To help relate the mean switching rate to the
switching current distribution width, we note that for a fixed
$T_{\text{b}}$, the shallower the slope of $\tau^{-1}(I,T_{\text{b}})$
the wider the corresponds distribution. The two main features of the
experimental data plotted in Fig.~\ref{Fig:MSR} are as $T_{\text{b}}$
decreases (1) the mean switching rate decreases ($\langle
I_{\text{sw}}\ \rangle$ increases) and (2) the slope of $\tau^{-1}(I)$
decreases ($I_{\text{sw}}$ distribution width becomes wider).

Two different fits to the same set of experimental data are shown in
Fig.~\ref{Fig:MSR}. The fit shown in the top panel includes TAPS only,
whilst the one in the bottom panel uses the fitting parameters from
the top panel but also includes QPS. The fits were obtained using the
fast switching-rate calculation routine described in
Appendix~\ref{App:Fast}. The tuning parameters that were obtained from
the fit are listed in Table~\ref{Table:Parameters} and fall into two
categories. The first category is composed of the geometric model
parameters, such as the wire length, whilst the second category
controls the ``input functions,'' i.e., the heat capacity, the heat
conductivity, and the phase-slip rate. The expressions for these input
functions are given in Appendix~\ref{App:Parameters}. We note that in
obtaining these fits we verified that the fitting parameters that we
used were consistent with the high-temperature $R(T)$
data~\cite{Sahu2008b}.

The TAPS-only fit (top panel of Fig.~\ref{Fig:MSR}) works well at
temperatures above $1\,\text{K}$. In this regime the theory is able to
quantitatively explain the observed rise in mean switching current
($\langle I_{\text{sw}}\rangle$) with decreasing temperature, as well
as the peculiar increase the $I_{\text{sw}}$ distribution width with
decreasing temperatures.

\subsection{Single-slip-to-switch regime}

In general, as the temperature is lowered and the bias current
increased, the wire tends to enter the single-slip-to-switch regime,
as indicated in Fig.~\ref{Fig:Map}. This regime roughly corresponds to
the region of the $(I,T)$ plane where a single phase slip heats up the
wire to $T_{c}(I)$, and thus the boundaries of this regime are
primarily determined by the heat capacity of the wire. Within this
regime, switching-current statistics correspond directly to the
phase-slip statistics.

Theoretical fitting indicates that at temperatures below $\sim
1\,\text{K}$ the wire enters the single-phase-slip-to-switch
regime. This regime is indicated by the switch of the theory curves
from solid red lines to dashed blue lines in Fig.~\ref{Fig:MSR}. In
the absence of quantum phase slips, in this regime the $I_{\text{sw}}$
distribution width should follow a more conventional behavior, and
decreases with temperature. This corresponds to the increase in the
slope of the mean switching rate curves with decreasing temperature in
the single-slip-to-switch regime (see the top panel of
Fig.~\ref{Fig:MSR}).

However, experimentally the distribution width seems to increase monotonically
as the temperature is lowered, even in the single-slip-to-switch regime. This
behavior suggests that there is an excess of phase slips at low temperatures.

\subsection{Quantum phase-slip scenarios}

We expect that at low temperatures quantum phase-slips will contribute
strongly to the switching rate. We model the presence of QPS by adding
their rate to the rate of TAPS, so as to obtain the total phase-slip
rate which goes into our model:
\[
\Gamma_{\text{total}}(I,T)=\Gamma_{\text{TAPS}}(I,T)+\Gamma_{\text{QPS}%
}(I,T).
\]
To model the QPS rate, $\Gamma_{\text{QPS}}(I,T)$, we replace $k_B T
\rightarrow k_B (T_{\ast}+T_{1}T)$ in $\Gamma_{\text{TAPS}}(I,T)$ (see
Appendix~\ref{App:Gamma}). Here, $T_{\ast}$ and $T_{1}$ are both
treated as fitting parameters. Letting $T_{1}$ be nonzero does
somewhat improve the quality of our fits.

We can envision several scenarios for the effect of QPS on the
switching current distributions, and these are summarized in
Fig.~\ref{Fig:distributionsVsT}. In the absence of QPS, upon lowering
the temperatures, once the single-slip-to-switch regime is reached the
distribution width will start decreasing with temperature. This type
of behavior is demonstrated by the $T_{\ast}=0$ line in
Fig.~\ref{Fig:distributionsVsT}. However, in the presence of QPS, the
distribution width is expected to saturate at low temperature, with
the saturation value controlled by $T_{\ast}$. If, upon cooling, the
single-slip-to-switch regime is reached before the temperature reaches
$T_{\ast}$, we expect the distribution width to first increase and
then decrease before saturating with decreasing temperature (cf.~the
$T_{\ast}=0.5\,\text{K}$ curve in Fig.~\ref{Fig:distributionsVsT}). On
the other hand, if $T_{\ast}$ is reached before the
single-slip-to-switch regime is reached, we expect the distribution
width to increase monotonically with decreasing temperature (cf.~the
$T_{\ast }=1.0\,\text{K},\,1.5\,\text{K}$ curves in
Fig.~\ref{Fig:distributionsVsT}).

To include QPS in our fitting, we started with parameters obtained by
fitting the mean switching rate curves at high temperatures
($T>1\,\text{K}$), as described in the previous subsection (i.e.~see
top panel of Fig.~\ref{Fig:MSR}).  Next, we optimized $T^{\ast}$ and
$T_{1}$ to obtain the best possible fit to the mean switching rate curves
at low temperatures as well. The optimal values thus obtained were
$T_{\ast}=0.726\,\text{K}$ and $T_{1}=0.4$, which corresponds to the
fit shown in the bottom panel of Fig.~\ref{Fig:MSR} and the curve
labeled \textquotedblleft best fit\textquotedblright\ in
Fig.~\ref{Fig:distributionsVsT}.

To fit the experimental data, we must be able to simultaneously match
both the mean and the standard deviation of the switching current
distribution as a function of temperature. However, we have not been
able to get quantitative agreement with both of these quantities,
simultaneously, in the low temperature regime. The parameter values of
$T_{\ast}=0.726\,\text{K}$ and $T_{1}=0.4$ result in a good fit of the
mean but not the standard deviation (see
Figs.~\ref{Fig:distributionsVsT} and \ref{Fig:MSR}), whilst the
values of $T_{\ast}=1.5\,\text{K}$ and $T_{1}=0$ result in a good fit
of the standard deviation (see Fig.~\ref{Fig:distributionsVsT}) but
not the mean (not shown).

We conclude this section by noting that, for the nanowire that we
fitted, our fitting seems to favor the QPS scenario where $T_{\ast}$
is higher than the temperature corresponding to the onset of the
single-slip-to-switch regime.

\section{Concluding Remarks}
\label{Sec:Conclusions}

In conclusion, we have developed a quantitative theory of stochastic
switching from the superconducting to the normal state in hysteretic
superconducting nanowires.  Our theory describes the dynamics of the
heating of nanowires by random-in-time phase-slip events, and the
cooling of the wire by heat conduction into the leads. In general, a
train of phase slips, sufficiently closely spaced in time can cause
the nanowire to overheat and switch from the low-temperature
(superconducting) branch to the high-temperature (normal) branch.

The main achievement of our theory is that it quantitatively describes
the unexpected increase in the switching current distribution with
decreasing temperature, as is observed in experiments.

Our theory also predicts that there is a single-slip-to-switch regime
at low temperatures. In this regime, a single phase slip always
triggers a switching event; thus, by studying the switching statistics
one has direct access to the phase slips statistics. Typically,
phase-slip properties have been studied by linear-response
measurements, which are only feasible when the nanowires have a
measurable resistance, i.e., at high temperatures. The
single-slip-to-switch regime is interesting because it occurs at low
temperatures, and thus it is a complementary tool with which to study
the properties of phase slips.

Finally, the monotonic increase of the switching current distribution
width with decreasing temperature, even in the single-slip-to-switch
regime, seems to indicate a severe excess of phase slips over the
predictions of the Langer-Ambegaokar McCumber-Halperin model of
thermally activated phase slips. It seems at the very least plausible
that the quantum tunneling of the superconducting order parameter
(i.e., QPS) is the mechanism that serves to meet this
excess~\cite{Sahu2008}.

\acknowledgments{ It is our pleasure to acknowledge invaluable
  discussions with T.-C. Wei, M.-H. Bae, A. Rogachev, and B. K. Clark.

  This material is based upon work supported by the
  U.S. Department of Energy, Division of Material Sciences under Award
  No.~DE-FG02-07ER46453, through the Frederick Seitz Materials
  Research Laboratory at the University of Illinois at
  Urbana-Champaign~(DP, MS, AB, PMG), and by the U.S. National Science
  Foundation under Award No.~DMR 0605813~(NS). PMG thanks for its
  hospitality the Aspen Center for Physics, where part of this
  research was carried out.  }

\appendix
\begin{widetext}
\section{Input functions and parameters}
\label{App:Parameters}
In this appendix we catalog the models for the phase-slip rate, heat
capacity, and thermal conductivity that go into the stochastic heat
equation~(\ref{heat-cond-eq}).
\subsection{Phase-slip rate}
\label{App:Gamma}
We begin by considering the phase-slip rate. For the TAPS rate,
$\Gamma_\text{TAPS}$, we have used the LAMH model, including the
nonlinear current response:
\begin{align}
\Gamma_\text{TAPS}(I,T)&=\Gamma_{-,\text{TAPS}}(I,T)-\Gamma_{+,\text{TAPS}}(I,T)
\label{Eq:f}, \\
\Gamma_{\pm,\text{TAPS}}(I,T)&=\Omega_{\pm}(I,T)\exp\left(-\frac{\Delta F_{\pm}(I,T)}{k_B T}\right), \label{Eq:GammaTAPS}
\end{align}
where $+$ or $-$ indicate whether the phase slip results in current
rise or drop, respectively. The phase-slip barriers $\Delta
F_\pm$ at bias current $I$ and temperature $T$ are given by
\begin{align}
\Delta F_{-}(T)&=C_1(T)\left(\frac{8}{3} \sqrt{2} \sqrt{1-3 k^2} - 8 k (1-k^2) \arctan\frac{\sqrt{1-3 k^2}}{\sqrt{2} k}\right),\\
\Delta F_{+}(T)&=C_1(T)\left(\frac{8}{3} \sqrt{2} \sqrt{1-3 k^2} + 8 k (1-k^2) \left[\pi-\arctan\frac{\sqrt{1-3 k^2}}{\sqrt{2} k}\right]\right),\\
C_1(T)&=\frac{3\sqrt{3}}{8} \frac{\hbar}{2 e} I_c(T),
\end{align}
where the phase gradient $k$ at current $I$ is the real solution of the equation
\begin{align}
\frac{I}{I_c(T)}=k(1-k^2),
\end{align}
and the temperature-dependent critical current $I_c(T)$~\cite{Bardeen1962}
is expressed in terms of the wire length $L$, the critical temperature
$T_c$, the zero-temperature coherence length $\xi_0$, and the normal-state
resistance of the wire $R_n$~\cite{TinkhamL2002}, via
\begin{align}
I_C(T)=(92\,\mu\text{A}) \frac{L T_c}{R_n \xi_0}
\left(1-\left(\frac{T}{T_c}\right)^2\right)^{3/2}.
\end{align}
We approximate the prefactor $\Omega_\pm(I,T)$ in Eq.~(\ref{Eq:GammaTAPS}) via
\begin{align}
\Omega(T)&=\frac{\sqrt{3}}{2 \pi^{3/2}}\frac{L}{\xi(T) \, \tau(T)}\left(\frac{\Delta F(I=0,T)}{k_B T}\right)^{1/2}\label{Eq:Omega}, \\
\Omega_{-}(I,T)&=(1-\sqrt{3} k)^{15/4} \, (1+k^2/4) \, \Omega(T),\\
\Omega_{+}(I,T)&=\Omega(T).
\end{align}
In the presence of a bias current $I$, the ``+'' phase slips are
exponentially more rare than the ``-'' phase slips. Therefore, we keep
the current-dependent corrections to the prefactor for the ``-'' phase slips,
but not for the ``+'' phase slips. Thus, we obtain an approximation that
works in both the linear-response regime, where the current correction is
irrelevant, and in the high-bias regime, where ``+'' phase slips are
rare. We estimate the temperature-dependent coherence length and the
Ginzburg-Landau relaxation time via
\begin{align}
\xi(T)&=\xi_0 \frac{\sqrt{1-(T/T_c)^4}}{1-(T/T_c)^2},\\
\tau(T)&=\frac{\pi \hbar}{8 k_B (T_c-T)}.
\end{align}
Thus, we can express the phase-slip rate via the physical parameters
$L$, $R_n$, $\xi_0$, $T_c$.  To obtain the quantum phase-slip rate, we
replace $\Delta F/k_B T$ in in Eqs.~(\ref{Eq:GammaTAPS}) and
Eq.~(\ref{Eq:Omega}) by $\Delta F/k_B T_\text{eff}$ where
$T_\text{eff}\equiv (T_*+T_1 T)$ is the effective temperature. $T_*$
and $T_1$ are treated as fitting parameters, $T^*$ being the
low-temperature limiting value of $T_\text{eff}$.

\subsection{Heat capacity and thermal conductivity}
Unfortunately, we know of no direct experimental data on the heat
capacity and thermal conductivity of current-carrying superconducting
nanowires. The diameter of the wires used in experiments is comparable
to $\xi_0$. Thus, the thermodynamic properties of these wires should lie
somewhere between those of a bulk superconductor and a normal
metal. Therefore, for the purposes of computing the thermodynamic
functions, we model the wire as being composed of a BCS superconducting
wire of cross-sectional area $A_1$ in parallel to a normal-metal wire of
cross-sectional area $A_2$.
The BCS and Fermi liquid expressions for heat capacity~\cite{Tinkham} are
\begin{align}
C_{v,\text{BCS}}(\Theta)&=-\frac{2 N_{0}}{\Theta} \int E_{k}\frac{df_{k}}{d(\beta E_{k})}\left(
E_{k}+\beta\frac{dE_{k}}{d\beta}\right)  d\xi_{k},\\
C_{v,\text{FL}}(\Theta)&=\frac{2}{3} \pi^2 N_0 k_B^2  \Theta,
\end{align}
where $\beta=1/k_B \Theta$,
$E_{k}=\sqrt{\xi_{k}^{2}+\Delta^{2}(\Theta)}$, $f_k$ is the Fermi
function, and $\Delta(\Theta)$ is obtained from the BCS gap
equation. Thus, the total heat capacity of the wire $C_v$ is given by
\begin{align}
C_v=\frac{A_1 C_{v,\text{BCS}} + A_2 C_{v,\text{FL}}}{A_1+A_2}.
\label{Eq:Cv}
\end{align}
Similarly, the dirty-limit BCS~\cite{BardeenRT1959, Kopnin2001} and Fermi-liquid
expressions for thermal conductivity are
\begin{align}
K_{s,\text{BCS}}(\Theta)&=2N_{0}D\int_{\Delta}^{\infty}\frac{\text{sech}^{2}\left[
\epsilon/2k_{B}\Theta\right]
}{2k_{B}\Theta}\frac{\epsilon^{2}} {k_{B}\Theta}d\epsilon,\\
K_{s,\text{FL}}(\Theta)&=\frac{L_{0}\Theta
L}{AR_{n}},
\end{align}
where $D$ is the diffusion constant (for MoGe
$D\sim1\,\text{cm}^{2}/\text{s}$~\cite{Graybeal1985}), and
$L_{0}=\pi^{2}k_{B}^{2}/3e^{2}$. The total thermal conductivity $K_s$ is,
correspondingly, given by
\begin{align}
K_s=\frac{A_1 K_{s,\text{BCS}} + A_2 K_{s,\text{FL}}}{A_1+A_2}.
\label{Eq:Ks}
\end{align}
The fitting parameters describing the heat capacity and thermal conductivity
are the cross-sectional areas $A_1$ and $A_2$, and $T_c$ of the nanowire.
\end{widetext}

\section{Fitting procedure}

\begin{table*}
\begin{tabular}{|c|c|l|l|}
\hline
type & parameter name & symbol & value \\
\hline
geometric & wire length & $L$ & $110\,\text{nm}$ \\
geometric & central segment length & $L_1$ & $110\,\text{nm}$\\
geometric & end segment length & $L_2$ & $27.5\,\text{nm}$ \\
input function & transition temperature & $T_c$ & $3.872\,\text{K}$ \\
input function & zero temperature transition length & $\xi_0$ & $5\,\text{nm}$ \\
input function & QPS effective temperature & $T_\text{eff}=T^*+T_1 T$ & 
$0.726 \,\text{K} + 0.4\, T $ \\
input function & effective superconducting cross-sectional area & $A_1$ 
& $320.4 \,\text{nm}^2$ \\
input function & effective normal cross-sectional area & $A_2$ & $19.0 \,\text{nm}^2$\\
input function & normal state resistance & $R_n$ & $2666\,\Omega$ \\
\hline
\end{tabular}
\caption{Parameters used in the switching rate model.  Parameters fall
  into two categories: geometric parameters and input function
  parameters. The former define the simplified model of the wire,
  while the latter define the phase slip rate, heat capacity and heat
  conductivity. Not all of these are used as tuning parameters, as
  $R_n$ and $L$ can be measured directly, and the fitting is only
  weakly effected by $L_1$ and $L_2$. Fitting parameters for a 
  representative experimental sample (corresponding to 
  Figs.~\ref{Fig:distributions}, \ref{Fig:Ic}, \ref{Fig:distributionsVsT}, 
  \ref{Fig:MSR}) 
  are displayed in the right hand column.}
\label{Table:Parameters}
\end{table*}

The main goal of the fitting procedure is to fit the switching-rate
data.  However, in addition to the mean switching rates at high bias
currents and low temperatures, we also have data on the
linear-response resistivity in the high-temperature regime. (The linear
response resistivity becomes too small to measure below
$T\sim1.9\,\text{K}$ for our wires.) The fitting is performed in two
steps.  In the first step, we fit the high-temperature linear-response
data. In the second step, we use the parameter values from the first
step as a starting point in fitting the switching-rate
data. Table~\ref{Table:Parameters} lists the parameters that go into
our model; the procedure for determining them is explained below.

We fit the high-temperature linear-response by conductivity following the usual
procedure~\cite{TinkhamL2002}. In this procedure, $L$ and $R_{n}$ are obtained
from microscopy and electrical measurements of the wire resistance above
$T_{c}$. We fit the $R(T)$ data using
\[
R=\lim_{I\rightarrow0}\frac{V}{I}=\lim_{I\rightarrow0}\frac{1}{I} \, \Phi_{0} \, \Gamma_{\text{TAPS}}(I,T)
\]
to obtain $T_{c}$ and $\xi_{0}$.

Next, we use the values of $L$, $R_{n}$, $\xi_{0}$, and $T_{c}$,
obtained in the first step, as a starting point in fitting of the mean
switching-rate data.  In this step, we tune $A_{1}$, $A_{2}$, $T_{c}$,
$\xi_{0}$, $T^{*}$, and $T_{1}$ simultaneously to obtain the best
possible fit over the entire current and temperature range. During
this procedure, we set $L_{1}=L$ and $L_{2}=L/4$.  We find that
variation of $L_{1}$ and $L_{2}$ does not significantly effect the fit,
and thus we exclude them from the already-extensive list of fitting
parameters.

Finally, we verify that the fitting parameters obtained from fitting
the mean switching-time data are consistent with the high-temperature
linear-response data.

\section{Fast mean first passage time calculation}

\label{App:Fast} 

\begin{figure}
  \includegraphics[width=8cm]{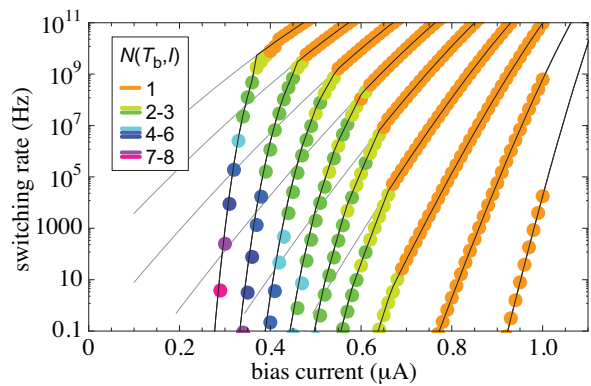}
  \caption{To compare approximate (thick solid lines) and exact
    (colored dots) methods for solving for the mean switching time
    delay-differential equation, we plot the mean switching rate
    $\tau^{-1}(T_{\text{b}},I)$ as a function of bias current $I$ for
    various bath temperatures $T_{\text{b}}%
    =\{1.9,\,1.7,\,1.5,\,1.3,\,1.1,\,0.9,\,0.7,\,0.5,\,0.3\}$ (from
    left to right)~\cite{FN:parameters}. The thin lines correspond to
    the phase-slip rate, and are shown for comparison. The parameter
    $k$ in Eq.~[\ref{Eq:defk}] was set to $0.5$.}
\label{Fig:ExactVsPh}
\end{figure}

In this subsection, we develop an approximation for
computing the mean switching time very quickly. This approximation
models the formation of a phase-slip train, and is useful for fitting
experimental data, where it is important to compute the mean switching
time for a lot of points $(T_{\text{b}}, \, I)$quickly.

In constructing this approximation we make several assumptions. First, we
assume that the phase-slip trains are dilute in time, i.e., the wire spends
most of its time at the temperature $T_{\text{b}}$, but very rarely there are
trains of phase slips that heat up the wire. These trains are not overlapping,
i.e., each train either leads to thermal run-away (a successful train) or the
wire cools back down to $T_{\text{b}}$ (an unsuccessful train). The train is
considered successful if the wire temperature exceeds $T^{\ast}$, as defined in
Section~\ref{Sec:LangevinProperties}.

In order to compute the switching rate, we must compute the
probability for the formation of a successful phase-slip train
$S(T_{\text{b}},I)$ and multiply it by the phase-slip rate
$\Gamma(T_{\text{b}},I)$, which corresponds to the rate of formation
of the first phase slip in a train. Thus the switching rate is given
by
\begin{align}
\tau^{-1}(T_\text{b},I)=S(T_\text{b}, I) \Gamma(T_\text{b},I).
\label{Eq:tauPh}
\end{align}
At this point, we make the additional assumption that the probability
to form a successful train can be computed phenomenologically, as
follows. Consider a phase slip in a wire that is at temperature
$T$. Immediately after a phase slip, the wire has temperature
$T+\eta(T,I)$, but it is also cooling at the rate
$r_{c}=(T-T_{\text{b}}) \, \alpha(T,T_{\text{b}})$. If the phase
slip-train is to continue, there must be another phase slip within a
time $\sim r_{c}^{-1}$; otherwise the wire would cool to the bath
temperature $T_{\text{b}}$ and the phase-slip train would be
unsuccessful. Applying this procedure to a chain of phase slips, we
find that $S(T_\text{b},I)$, the probability to construct a successful
phase-slip train, is given by
\begin{align}
S(T_\text{b},I) \sim \prod_{i=1}^N \frac{\Gamma(T_i, I)}{(T_i-T_\text{b}) \alpha(T_i, T_\text{b})}.
\label{Eq:S1}
\end{align}

The mean switching rate, computed using the phenomenological model for
the probability to form a successful phase slip train given by
Eq.~(\ref{Eq:S1}), turns out to be too crude to give results that are
quantitatively accurate, although, qualitatively, the exact results
obtained by solving Eq.~(\ref{Eq:tau}) are well reproduced. In order
to improve accuracy, we take into account the fact that the phase-slip
rate drops as the wire cools, and also introduce the tunable parameter
$k$, which characterizes how much the wire is allowed to cool before a
phase-slip train is considered to be unsuccessful. Consider a wire at
temperature $T_{i}$. In the absence of phase slips, we approximate
the equation for the evolution of the wire temperature by
\begin{align}
\partial_t T(t)=-(T-T_\text{b}) d_i,
\end{align}
where $T(0)=T_{i}$, and $d_{i}=C_{1}(T_{i},T_{\text{b}})$. To parametrize the
failure of a phase-slip train, we assume that a phase-slip train is
unsuccessful if the temperature of the wire reaches the value
\begin{align}
T_{i,\text{fail}}=(1-k) T_i + k T_{i-1},
\label{Eq:defk}
\end{align}
where $k$ is a tunable parameter of order unity that should be chosen to
minimize the difference between the exact [i.e., obtained from solutions of
Eq.~(\ref{Eq:tau})] and the phenomenological switching rates. Having defined
$T_{i,\text{fail}}$, we can define the time to reach it via
\begin{align}
t_{i,\text{fail}}=\frac{1}{d_i} \log \left(
\frac{T_{i,\text{fail}}-T_\text{b}}{T_i-T_\text{b}} \right).
\end{align}
Finally, taking into account the change in the phase-slip rate as the wire
cools, as well as $t_{i,\text{fail}}$, we modify Eq.~(\ref{Eq:S1}) to read
\begin{align}
S(T_\text{b},I) \sim \prod_{i=1}^N \int_0^{t_{i,\text{fail}}} dt\,
\Gamma \left( T_\text{b}+(T_i-T_\text{b}) e^{-d_i t}, I \right). 
\label{Eq:S2}
\end{align}

In Fig.~\ref{Fig:ExactVsPh} we compare the approximate switching rates
obtained from Eq.~(\ref{Eq:tauPh}) by using the phenomenological
approximation Eq.~(\ref{Eq:S2}) to the exact switching rate obtained
from Eq.~(\ref{Eq:tau}). We see that the phenomenological
approximation is quantitatively very close to the exact switching
rate. Therefore, to performing fits on experimental data we, in fact,
use this phenomenological approximation for the switching rate, as it
can be computed much faster.


\begin{thebibliography}{99}
\bibitem{FN:setting}In settings in which it is instead the current
  carried by the system that is externally controlled, $\Psi(z)$ does
  undergo a transition, but the boundary values of $\Psi(z)$ evolve in
  concert with the bulk values to produce a final state that is
  identical to the initial one.  The imposed current is carried
  through the nanowires almost exclusively as supercurrent, except in
  the (temporal and spatial) neighborhood of the phase-slip event,
  when and where it is primarily carried as normal current.
\end{thebibliography}
\end{document}